\documentclass[12pt]{article}
\usepackage{amssymb,amsfonts,latexsym}

\begin{document}

\def\ns{\normalsize}

\renewcommand{\thesection}
             {\Roman{section}}

\def\op{\sharp}
\def\pop{Q}
\def\Al{\Xi_0}

\def\A{{\cal A}}
\def\Aa{{\mathfrak A}}
\def\AA{A^\sharp}
\def\B{{\bf B}}
\def\C{{\bf C}}
\def\H{{\bf H}}
\def\I{{\mathfrak I}}
\def\IPa{I_{P_a}}
\def\IQa{I_{Q_a}}
\def\IP{I_{P_0}}
\def\IQ{I_{Q_0}}
\def\IbPa{I_{\bar{P}_a}}
\def\IbQa{I_{\bar{Q}_a}}
\def\IbP{I_{\bar{P}_0}}
\def\IbQ{I_{\bar{Q}_0}}
\def\Jj{{\cal J}}
\def\K{A_a}
\def\Mm{{\mathfrak M}}
\def\N{{\bf N}}
\def\R{{\bf R}}
\def\S{{\bf S}}
\def\Z{{\bf Z}}
\def\T{{\cal T}}
\def\TT{{\bf T}}
\def\Q{{\cal Q}}
\def\bbeta{\bar{{\mathfrak p}}}
\def\alph{{\mathfrak p}}
\def\PV{\pi_V}
\def\PVc{\bar{\pi}_V}
\def\sig{s}
\def\tPhi{\tilde{\Phi}}
\def\cL{{\cal L}}
\def\cF{{\mathfrak F}}
\def\tJ{\tilde{J}}

\def\1{{\bf 1}}
\def\phys{{\mathfrak M}_{phys}}
\def\grad{\nabla_g}
\def\crit{{\mathfrak C}}
\def\cg{\crit_{gen}}
\def\ker{{\rm ker}}

\def\prf{{\noindent\bf Proof.}$\;$}
\def\qed{\vrule height .6em width .6em depth 0pt\bigbreak}

\def\eqnn{\begin{eqnarray*}}
\def\eeqnn{\end{eqnarray*}}
\def\eqn{\begin{eqnarray}}
\def\eeqn{\end{eqnarray}}
\def\nbf{}
\def\nit{}

\def\bc{\begin{center}}
\def\ec{\end{center}}

\newtheorem{thm}{Theorem}[section]
\newtheorem{dfi}{Definition}[section]
\newtheorem{prp}{Proposition}[section]
\newtheorem{hyp}{Hypothesis}[section]
\newtheorem{lm}{Lemma}[section]
\newtheorem{cor}{Corollary}[section]
\newtheorem{conj}{Conjecture}[section]

$\;$
\\
\\
\\
\\
\begin{center}
{\huge\bf NON-HOLONOMY, CRITICAL
\\

MANIFOLDS AND STABILITY IN
\\

CONSTRAINED HAMILTONIAN
\\

SYSTEMS \\ $\;$}
\end{center}
\begin{center}
Thomas Chen\footnotemark
\\Theoretische Physik\\ETH H\"onggerberg\\8093 Z\"urich\\
Switzerland\\
chen@itp.phys.ethz.ch
\end{center}

$\;$
\\
\noindent{\bf Abstract}\\
We approach the analysis of dynamical and
geometrical properties of nonholonomic mechanical systems
from the discussion of
a more general class of auxiliary constrained Hamiltonian systems.
The latter is constructed in a manner that it comprises the
mechanical system as a
dynamical subsystem, which is confined to an invariant manifold.
In certain aspects, the embedding system can be more easily analyzed than
the mechanical system.
We discuss the geometry and topology of the critical set of either
system  in the generic case, and prove results closely related to the strong
Morse-Bott, and Conley-Zehnder
inequalities.
Furthermore, we consider qualitative
issues about the stability of motion in
the vicinity of the critical set. Relations
to sub-Riemannian geometry are pointed out, and possible
implications of our results for engineering problems are sketched.
\\
\\
\\
\\
\\
\\

\footnotetext{This is a shortened and modified version of the author's
PhD thesis in mechanical
engineering at ETH Z\"urich, from January 1999 (ETH-Diss 13017).
The work was carried out at the institute of mechanics (now IMES),
ETH Z\"urich,  before
October 1998.}

\thispagestyle{empty}
\pagebreak

\setcounter{page}{1}

\section*{Introduction}

This text is a shortened and modified version of the author's PhD
thesis in mechanical engineering at ETH Z\"urich,
(ETH-Diss 13017, January 1999).
Its subject matter is located in the area of theoretical mechanics, and
discusses a particular class of constrained Hamiltonian systems,
which contains the subclass of nonholonomic mechanical systems.
We will be mainly
concerned with geometrical and
global topological issues about the critical manifolds,
the stability of equilibria, aspects about numerical stability,
and technical applications.

An area  
of the engineering sciences, where nonholonomic mechanical
systems are
typically encountered, is the field of multibody systems simulation.
There is a multitude of different approaches to 
the description and analysis of such systems, stemming from
various subareas of application.
The geometrical approach given here is influenced by the 
work of R. Weber \cite{We} on Hamiltonian systems with
nonholonomic constraints,
and very essentially by the formalism developed by H. Brauchli based on
an orthoprojector calculus, \cite{Bra,Bra2,SoBr}.
A construction
for the Lagrangian case, which is closely related to
what will be presented in section
3,  has been
given by Cardin and Favretti, \cite{CaFa}. A different approach
in the Hamiltonian picture is dealt with in the work of 
J. Van der Schaft and B.M. Maschke, \cite{vdSMa}.
A geometrical theory of nonholonomic systems with a strong
influence of network theory has been developed by
H. Yoshimura and T. Kawase, \cite{YoKa}.
The geometrical structure of nonholonomic systems
with symmetries and the associated reduction theory, as well as
aspects of their stability theory has been at the focus in
the important works of J.M. Marsden and his collaborators,
see for instance \cite{BlKrMaMu,KoMa,ZeBlMa}.

In section one, we introduce a particular class of constrained
Hamiltonian systems.
Given a symplectic manifold $(M,\omega)$
and
a symplectic distribution $V$, we focus on the flow $\tPhi_t$
generated by the component $X_H^V$
of the Hamiltonian vector field $X_H$ in $V$.

Section two addresses the geometry and global topology of the
critical set $\crit$ of the constrained
Hamiltonian system.
The main technical tool used for this purpose is a gradient-like flow
$\phi_t$,
whose critical set $\crit$ is identical to that of $\tPhi_t$.
Assuming that the Hamiltonian $H:M\rightarrow\R$ is a Morse
function, it is proved that generically, $\crit$ is a normal hyperbolic
submanifold of $M$.
The generalization of Morse theory developed by C. Conley and
E. Zehnder is used to prove a topological formula for closed, compact $\crit$,
that is closely related to the Morse-Bott inequalities.
A second proof is given, based on the use of the Morse-Witten
complex in Morse theory, to further clarify certain structural aspects.

In section three, we address questions about the stability
of the constrained Hamiltonian systems in discussion, and conjecture
a stability criterion for the marginally stable case. Relations
to sub-Riemannian geometry are pointed out.

In section four, we introduce
Hamiltonian mechanical systems with Pfaffian constraints, which
represent the case of highest technical relevance.
It is shown that the physical orbits of such a system
are confined to a smooth submanifold $\phys\subset T^*Q$.
We construct an
auxiliary constrained Hamiltonian system of the type introduced
in section one, which  exhibits $\phys$ as an invariant manifold,
on which it generates the physical flow.
Furthermore, we study the global topology of
the critical manifold of the physical constrained system, and
analyze
the stability of equilibria of such systems.
Finally,
we propose a method to
numerically construct the generic connectivity components
of the critical manifold.
\\

\section*{\bf Acknowledgements}

I warmly thank my thesis advisor Prof. H. Brauchli for
suggesting this area of problems and
the possibility to carry out this work, as well as
for original and inspiring ideas.
I am profoundly grateful to Prof. E. Zehnder for very generously
dedicating his precious time to a great number of
discussions that were quintessential for my proper understanding of the
topics presented
in the sequel, and for accepting the coadvisorship.
I am deeply grateful to Prof. J. Fr\"ohlich
for his advice and support.
I also thank Prof. M.B. Sayir
and the IMES (formerly the institute of mechanics) at
ETH Z\"urich for an excellent infrastructure, 
where this work was done before October 1998.

I am also much indebted to M. Clerici for his insights and many discussions,
to M. von Wattenwyl for discussions, and to
Prof. M. Sofer for his
help at early stages of this work. It is a pleasure to thank Prof. H.
Yoshimura for highly interesting and inspiring discussions, and to Prof.
O. O'Reilly for his friendliness
and hospitality during a conference trip to California.

\tableofcontents

\pagebreak

\section{A NONINTEGRABLE GENERALIZATION OF \\
DIRAC CONSTRAINTS}
\label{gendir}

\noindent{\bf Hamiltonian dynamics}.
Let $(M,\omega,H)$ be a Hamiltonian system. $M$ denotes a smooth,
symplectic $2n$-manifold with a $C^\infty$ symplectic structure
$\omega\in \Lambda^2(M)$.
$H\in C^\infty(M)$ is the Hamilton function, or
the Hamiltonian in brief. For $p=1,\dots,2n$,
$\Lambda^p(M)$ is the $C^\infty(M)$-module of $p$-forms on $M$.
The Hamiltonian vector field $X_H\in \Gamma(TM)$ is determined by
$$i_{X_H}=-dH\;,$$
where $i$ stands for interior multiplication by a vector field.
Given a smooth distribution $W\subset TM$,
$\Gamma(W)$ will denote the $C^\infty(M)$-module
of smooth sections of $W$.
The Hamiltonian flow is the 1-parameter group
$\Phi_t\in {\rm Diff}(M)$ generated by $X_{H}$, with $t\in \R$,
and $\Phi_0={\rm id}$.
Its orbits obey
\eqn\label{hameqmo}\partial_t\Phi_t(x)\;=\;X_H(\Phi_t(x))\;\eeqn
for all $x\in M$, and $t\in\R$.

$\omega$ induces a smooth, non-degenerate Poisson structure on $M$,
the $\R$-bilinear, antisymmetric pairing on $C^\infty(M)$ given by
\eqn\label{poisson}\lbrace f,g\rbrace=\omega(X_f,X_g)\;.\eeqn
It is a derivative in both of its arguments,
and satisfies the Jacobi identity
$$\lbrace f,\lbrace g,h\rbrace\rbrace+\lbrace h,\lbrace f,g\rbrace\rbrace+
\lbrace g,\lbrace h,f\rbrace\rbrace = 0\;,$$
thus $(C^\infty(M),\lbrace\cdot,\cdot\rbrace)$ is a Lie algebra.
Along every orbit of the Hamiltonian flow,
\eqn\label{Poisseqsmo}\partial_t f(\Phi_t(x))\;=\;
     \lbrace H, f\rbrace(\Phi_t(x))\eeqn
is satisfied for all $f\in C^\infty(M)$, and all $x\in M$, $t\in\R$.
\\

\noindent{\bf Dirac constraints}.
Let $j:M'\rightarrow M$ be an embedded, smooth,
$2k$-dimensional symplectic submanifold of $M$,
endowed with the pullback symplectic
structure $j^*\omega$. The induced
Poisson bracket on $M'$ is  defined by
$$\lbrace f,g\rbrace_D\;=\;
      (j^*\omega)(X_{\tilde{f}},X_{\tilde{g}})$$
for any pair of extensions $\tilde{f},\tilde{g}\in C^\infty(M)$ of
$f,g\in C^\infty(M')$, and is in
this context referred to as the Dirac bracket.

If $M'$ is locally characterized as the locus of common
zeros of some family of functions $G_i\in C^\infty(M)$, with
$i=1,\dots,2(n-k)$, it is possible to give the following
explicit construction of the Dirac bracket,  \cite{MaRa}.
Because $M'\subset M$ is symplectic, the $(n-k)^2$ quantities
$$D_{ij}\;:=\;\lbrace G_i,G_j\rbrace$$
define a smooth, matrix-valued function that
is invertible everywhere on $M'$.
The explicit formula for the Dirac bracket is
given by
\eqn\label{dirbra}\lbrace f,g\rbrace_D\;=\;
       \lbrace \tilde{f},\tilde{g}\rbrace\;-
      \;\lbrace \tilde{f},G_i\rbrace
      \;D^{ij}\;\lbrace G_j,\tilde{g}\rbrace\;,\eeqn
where $D^{ij}$ denotes the components of the inverse of
$[D_{ij}]$.
\\

\noindent{\bf A natural generalization}. We will next
mimick this construction in the following more general setup.

A distribution $V$ over the base manifold $M$ will be called symplectic
if $V_x$ is a symplectic subspace of $T_x M$ for all $x\in M$
with respect to $\omega_x$. Accordingly,
$V^\perp$ denotes
the distribution characterized by the property that
its fibres are the symplectic complements
of those of $V$ with respect to $\omega$, and is symplectic.
Smoothness of $V$ and $\omega$ implies smoothness
of  $V^\perp$.
If $V$ is symplectic, the Whitney sum bundle
$V\oplus V^\perp$ equals $TM$.
\\

Thus, let $V$ denote an
integrable, smooth, symplectic rank $2k$-distribution $V$ over $M$.
Then, any section $X\in\Gamma(TM)$ can be written as
$$X=X^V + X^{V^\perp}\;,$$
where $X^{V^{(\perp)}}$ are sections in $\Gamma(V^{(\perp)})$.
Moreover,
$\omega(X^V , X^{V^\perp})=0$, by definition of $V^\perp$.
Associated to this decomposition of vector fields into
components lying in
$V$ and $V^\perp$,
there exists an $\omega$-orthogonal tensor
$\PV:TM\longrightarrow TM$ with
$${\rm Ker}(\PV)=V^\perp\;\;\;\;,
    \;\;\;\;\PV(X)=X\;\;\;\;\forall X\in\Gamma(V)\;,$$
which
satisfies
$$\omega(\PV (X),Y)=\omega(X,\PV (Y))$$
for all $X,Y\in\Gamma(TM)$.
It will be referred to as the $\omega$-orthogonal
projector associated to $V$.

Given a local family of spanning vector fields
$Y_1,\dots,Y_{2k}$ for $V$, one can obtain an explicit formula for
$\PV$ in a construction similar to
(~\ref{dirbra}).
To this end, one uses the fact that symplecticness of $V$ is
equivalent to the condition of
invertibility of  the matrix $C_{ij}:=\omega(Y_i,Y_j)$.

\begin{lm}\label{Pconstr}
Let $C_{ij}:=\omega(Y_i,Y_j)$, and let $C^{kl}$ denote the components of
its inverse. Define the 1-forms $\theta_j(\cdot):=\omega(Y_j,\cdot)$.
Then, locally,
$\PV=C^{ij}Y_i\otimes\theta_j$.
\end{lm}

\prf
Clearly, the range of $C^{ij}Y_i\otimes\theta_j$ is $V$.
Furthermore,
$$(C^{ij}Y_i\otimes\theta_j)^2=C^{ij}Y_i\otimes\theta_j\;$$
because of $\theta_i(Y_j)=C_{ij}$, and $C_{ij}C^{jk}=\delta_i^l$.
To prove $\omega$-orthogonality for the asserted expression
for $\PV$, consider an arbitrary pair of vector fields $X,X'$, and
observe that
\eqnn \omega(\PV (X),X')&=&\omega(C^{kl}\theta_l(X)Y_k,X')\\
         &=&C^{kl}\theta_l(X) \omega(Y_k,X')\\
         &=&C^{kl}\theta_l(X)\theta_k(X')\\
         &=&-C^{lk}\theta_l(X)\theta_k(X')\\
         &=&-\omega(Y_l,X) C^{lk}\theta_k(X')  \\
         &=&\omega(X,C^{lk}\theta_k(X')Y_l )\\
         &=&\omega(X,\PV (X'))\;,\eeqnn
which establishes the claim. \qed

As a generalization of the usual Hamiltonian flow,
the following dynamical system can be naturally associated
to $(M,\omega,H,V)$.
The component of $X_H$ in $V$,
$$X_H^V\;:=\;\PV( X_H)\;\;\;\;\subset\;\Gamma(V)\;,$$
generates a 1-parameter group of diffeomorphisms  $\tPhi_t$.
The orbits of $\tPhi_t$ obey
\eqn\label{eqsofmo}\partial_t \tPhi_t(x)\;=\;X_H^V(\tPhi_t(x))\;,\eeqn
and are tangent to $V$ for all initial conditions $x\in M$.
Embeddings $\R\rightarrow M$ that are tangent to $V$ are
called $V$-horizontal. If the condition of integrability
imposed on $V$ is dropped,
this dynamical system will allow for the description of
nonholonomic mechanics.

Integrability of $V$ implies that $M$ is foliated
in terms of
$2k$-dimensional integral manifolds of $V$, which are, due to the
symplecticness of $V$,  symplectic submanifolds of $M$.
Therefore, every leaf $j:M'\rightarrow M$ is an embedded
invariant manifold under the action of $\tPhi_t$.

The induced dynamical system on any fixed leaf is equivalent to the
pullback Hamiltonian system $(M,j^*\omega,H\circ j)$
that has been considered before in the discussion of Dirac constraints.
In this sense, (~\ref{eqsofmo}) generalizes the notion of
Dirac constraints.
\\

\noindent{\bf Non-integrability}.
A new class of dynamical systems is obtained by discarding the
requirement of integrability on  $V$.

By definition, the distribution $V$ is non-integrable if there
exists a filtration
\eqn\label{filt}V_0\subset V_1\subset V_2\subset\cdots\subset V_r,\eeqn
inductively defined by $V_0=V$, and
$$V_i=V_{i-1}+[V_0,V_{i-1}]\;,$$
which is non-trivial, that is, $V_1\neq V_0$.
$[\cdot,\cdot]$ denotes the Lie bracket, and
the sequence $\lbrace V_i\rbrace_1^r$ is called the {\em flag} of $V$.
For a pair of smooth distributions $W,W'$, the distribution
$[W,W']$ is locally given by the linear
span of all Lie brackets that can be taken
between sections of $W$ and $W'$.
If the
fibre ranks of all $V_i$ are base point independent,
$V$ is called equiregular.
The smallest number $r(V)$ at which the flag of $V$
stabilizes, that is, for which
$$V_s=V_{r(V)}\;\;\;\;\forall s\geq r(V)\;,$$
is called the
degree of non-holonomy of $V$.
If $V_{r(V)}=TM$, $V$ is said to satisfy Chow's condition, or to be
totally non-holonomic.

\begin{prp}\label{Frob}(Frobenius condition)
The symplectic distribution $V$ is integrable if and only if
in local coordinates,
\eqn\cF^k_{ij}\;:=\;(\PV)^r_i\;(\PV)^s_j\;\left(\partial_r(\PV)^k_s-
      \partial_s(\PV)^k_r\right)\;=\;0\;\eeqn
holds everywhere in $M$.
\end{prp}

\prf
Clearly, $V$ is integrable if and only if
$\PVc\left([\PV (X),\PV (Y)]\right)=0$
is satisfied by all sections $X,Y$ of $TM$, which is equivalent
to $V_1=V$. The asserted
formula is the local coordinate representation of this
condition. \qed

\noindent{\bf Local coordinate representation}.
In a local Darboux chart with coordinates $x^i$,
the symplectic structure is represented by the matrix
\eqn\label{sympJ}{\cal J}\;=\;
      \left[\begin{array}{cc}0&\1_n\\-\1_n&0\end{array}\right]\;.\eeqn
Writing $x(t)$ for the coordinate components of $\tPhi_t(x_0)$
with $x_0\in M$, the equations of motion
(~\ref{eqsofmo}) are represented by
\eqn\partial_t x(t)&=&
     \left(P_x\;{\cal J}\;\partial_x H\right)(x(t))\nonumber\\
    &=& \left({\cal J}\; P_x^\dagger\;\partial_x H\right)(x(t))\;.
    \label{eqsmocoord}\eeqn
$P$ denotes the matrix of $\PV$, and $P^\dagger$ is
its transpose.
Due to lemma {~\ref{Pconstr}}, $P$ can be explicitly
determined once one picks
a local spanning family of vector fields for $V$.
\\

\noindent{\bf Almost K\"ahler structure.}
Let $g$ denote a Riemannian metric
on $M$.
An almost complex structure $J$ is a smooth bundle
isomorphism $J:TM\rightarrow TM$ with
$J^2=-\1$. Together with $g$, it defines a two form satisfying
\eqn\label{sksadness}\omega_{g,J}(X,Y)=g(JX,Y) \eeqn
for all sections $X$, $Y\in\Gamma(TM)$.
The Riemannian metric $g$ is hermitean if
$g(JX,JY)=g(X,Y)$, and {\em K\"ahler} if $\omega_{g,J}$
is closed. Thus, if  $\omega_{g,J}=\omega$ (the symplectic
structure), $g$ is K\"ahler. A triple $(g,J,\omega)$ of this
type is called compatible.

\begin{prp}
There is a compatible triple $(g,J,\omega)$, where $\omega$
is the symplectic structure, such that
$P$ is symmetric with respect to $g$, and $\PV JX=J\PV X$
for all $X\in\Gamma(TM)$.
\end{prp}

\prf
Let us pick a smooth Riemannian metric $\tilde{g}$ on $M$,
relative to which $P$ shall be symmetric.
For instance, one may choose an arbitrary Riemannian metric $g'$ on $M$,
and use it to define
$\tilde{g}(X,Y)\equiv g'(\PV X,\PV Y)+g'(\bar{\pi}_V X,\bar{\pi}_V Y) ,$
where $\bar{\pi}_V$ denotes the projection complementary to $\PV$.

Since $\omega$ is non-degenerate and smooth, there
is a non-degenerate, smooth bundle mapping $K$ which is defined by
\eqn\omega(X,Y)=\tilde{g}(K X,Y).\eeqn
It is skew symmetric with respect to $\tilde{g}$,
because the symplectic structure is antisymmetric, so the adjoint operator of
$K$ with respect to $\tilde{g}$ is $K^*=-K$.

The bundle map $K$ can be used to construct an almost symplectic structure.
Since $K^* K=-K^2$ is smooth, positive definite, non-degenerate and
symmetric,
there is a unique smooth, positive definite, symmetric bundle map $A$
defined by
$A^2=-K^2$,  which commutes with $K$.   This immediately implies
that the bundle map $J=KA^{-1}$ satisfies
$J^2=-\1$, so that it defines an almost complex structure.
This is the standard proof for the fact that every almost
symplectic manifold admits an almost complex structure, see for instance
\cite{Bry}.

Because $A$ is positive definite and symmetric, one can define a new metric
by $g(X,Y)\equiv\tilde{g}(AX,Y)$,
which obviously satisfies
\eqn\omega(X,Y)=g(JX,Y) .\eeqn
Moreover, this metric is hermitean, since
\eqnn g(JX,JY)&=&\tilde{g}(KX, A^{-1}K Y)\\
       &=& -\tilde{g}(X,K^2 A^{-1}Y) \\
       &=&\tilde{g}(X,AY)\\
       &=&g(X,Y) .\eeqnn
Because $\omega$ is closed, $g$ is not only hermitean, but even K\"ahler.

The projector $\PV$ is symmetric with respect to $g$, as one concludes from the
following consideration. The fact that
$P$ commutes with $K$ follows from
$$\tilde{g}(K\PV X,Y)=\omega(\PV X,Y)=\omega(X,\PV Y)=\tilde{g}(KX,\PV Y)=
      \tilde{g}(\PV KX,Y)$$
for all $X$ and $Y$, since $\PV$ is symmetric with
respect to $\tilde{g}$.
$P$ commutes with $K$, so also commutes with $K^2$;
thus it commutes with $A^2$.
The linear operator $A^2$ is positive definite and symmetric,
therefore commutativity with the symmetric operator $P$ implies that $P$ also
commutes
with $A$. It immediately follows that $P$ is symmetric with respect to $g$.

Finally, let us prove that $\PV$ commutes with $J$ on $\Gamma(TM)$.
To this end,
we observe that
$$g(J\PV X,Y)=\omega(\PV X,Y)=\omega(X,\PV Y)=g(JX,\PV Y)=g(\PV JX,Y)$$
holds for all $X$ and $Y$, as a result of which one concludes that $\PV J X=J\PV X$ is
satisfied for all $X$, as claimed. This implies that $J$ restricts to a bundle map
$J:V\rightarrow V$. \qed

If the $J$ is covariantly constant with respect to the
Levi-Civita connection associated to $g$, it
is a complex structure. If $\omega$ admits a compatible K\"ahler metric
$g$ together with a complex structure $J$, $(M,g,J)$ is a K\"ahler
manifold.
\\

\noindent{\bf Symmetries}.
Given a Lie group $G$, $(M,\omega,H)$ admits a
symplectic $G$-action
$\Psi:G\rightarrow {\rm Diff}(M)$, if
$$\Psi_h^*\;\omega\;=\;\omega\;\;\;,\;\;\;H\circ\Psi_h\;=\;H$$
is satisfied for all $h\in G$.
Accordingly, for the case of a constrained Hamiltonian system,
we will say that $(M,\omega,H,V)$ possesses a $G$-symmetry
if $\Psi_{h*} V=V$ holds for all $h\in G$.
\\

\noindent{\bf Generalized Dirac bracket}.
In the same spirit in which one defines the Poisson structure
induced by $\omega$, it is possible to introduce a smooth, $\R$-bilinear,
antisymmetric pairing on $C^\infty(M)$
that is associated to the triple $(M,\omega,V)$.
In fact
\eqn\label{Vbra}\lbrace f,g\rbrace_V\;:=\;\omega(\PV(X_f),\PV(X_g))\;\eeqn
is a straightforward generalization of  the Dirac and Poisson brackets.
Along orbits of $\tPhi_t$, one has
$$\partial_t f(\tPhi_t(x))\;=\;\lbrace H,f\rbrace_V(\tPhi_t(x))\;$$
for all $x\in M$,
in analogy to (~\ref{Poisseqsmo}).
The bracket (~\ref{Vbra}) does not satisfy the Jacobi identity
if $V$ is non-integrable, but it
satisfies a Jacobi identity on every
(symplectic) integral manifold if $V$ is integrable.
\\

\noindent{\bf Energy conservation and symplecticness}.
Let us finally mention two further important properties of the flow
$\tPhi_t\in {\rm Diff}(M)$, $t\in\R$, of the constrained system.

\begin{prp}
The energy $H$ is an integral of motion of the dynamical system
(~\ref{eqsofmo}).
\end{prp}

\prf
This follows from the antisymmetry of the generalized Dirac bracket,
which implies that $\partial_t H=\lbrace H,H\rbrace_V=0$. \qed

Finally, let us consider
\eqnn\partial_t\tPhi_t^{*}\omega&=&\tPhi_t^*\;\cL_{X_H^V}\omega\\
      &=&  \tPhi_t^*\left(d i_{X_H^V}\omega + i_{X_H^V}d\omega\right)\\
      &=& -\;\tPhi_t^*\;d\left(\;(P_V)^i_k \;\partial_i H \; dx^k\;\right)\\
      &=& -\;\tPhi_t^*\;\left(\;\partial_l(P_V)^i_k\;\partial_i H\;
             dx^l\wedge dx^k\;\right)\\
      &=& -\;\tPhi_t^*\;
           \left(\;\frac{1}{2}\;\left(\;\partial_l(P_V)^i_k -
           \partial_k (P_V)^i_l\;\right)\;
           \partial_i H\;dx^l \wedge dx^k\;\right)\;.\eeqnn
This result implies that the restriction of
$\partial_t\tPhi_t^*\omega$ to $X,Y\in\Gamma(V)$ satisfies
\eqnn\partial_t\tPhi_t^*\omega(X,Y)\;=\;
      -\;\tPhi_t^*\;
       \left(\;\frac{1}{2}\;
       \cF^i_{rs}\;\partial_i H\;X^r Y^s\;\right)
       \;,\eeqnn
where $\cF^i_{rs}$ is defined in lemma {~\ref{Frob}}, which
states that the the right hand side vanishes identically if and
only if $V$ is integrable.

Thus, if $V$ is integrable,
the restriction of $\tPhi_t^*\omega$ to $V\times V$
 equals its value for $t=0$, given by
$\omega(\PV(\cdot),\PV(\cdot))$. Hence, on every
integral manifold $j:M'\rightarrow M$ of $V$,
 $\tPhi_t$ is symplectic with respect
to the pullback symplectic structure $j^*\omega$.

\pagebreak

\section{\nbf THE GEOMETRY AND TOPOLOGY OF THE \\
CRITICAL MANIFOLD}

The critical set $\crit$ of a dynamical system is defined as
the set of equilibrium solutions.
The main purpose of this section is to study geometrical and
global topological properties of the critical sets $\crit$
exhibited by constrained
Hamiltonian systems of the type $(M,\omega,H,V)$.
An application of Sard's theorem
will demonstrate that generically, $\crit$ is a
smooth $2(n-k)$-dimensional submanifold $\cg\subset M$,
the critical manifold.

Moreover, it will be demonstrated that there exists a
precise correspondence between $\cg$ and the critical points of
$H$ if $H:M\rightarrow\R$ is picked to be a Morse
function.
In particular, we will prove that for compact $M$ without boundary,
the Poincar\'e polynomials of $M$ and those of the connectivity components of
$\cg$ are related in a way reminiscent of the Morse-Bott inequalities.

Two different proofs will in fact be presented.
The first proof is based on the generalization of Morse and
Morse-Smale theory \cite{Bo1,Hi,Mi,Sm} developed by C. Conley and
E. Zehnder in \cite{CoZe}, applied to an auxiliary gradient-like
system.
The second proof is based on the comparison of the
Morse-Witten complexes of $(M,H)$ and $(\cg,H|_{\cg})$.

The special case of
mechanical systems (where $M$ is noncompact) will be analyzed
in the last section.
\\

The following standard definitions
are necessary for the subsequent discussion.

\begin{dfi}
The zeros of the one form $dH$ are called  critical points of $H$,
and the value of $H$ at a critical point is called a critical value.
A level surface $\Sigma_{E}$ that contains a critical value $E$ of $H$
is called a critical level surface.
A level surface $\Sigma_E$ that contains no
critical points of $H$ is called regular, and consequently,
$E$ is then called a regular value.
\end{dfi}

\subsection{\nbf Generic properties of the critical set}

The critical set of the constrained Hamiltonian system
$(M,\omega,H,V)$ is given by
$$\crit\;=\;\left\lbrace x\in M\;|\;X_H^V(x)\;=\;0\right\rbrace\;
    \;\subset\;\;M\;.$$
The following theorem holds independently
of the fact whether $V$ is integrable or not.

\begin{thm}\label{Sardthm}
Generically, the critical set is a piecewise
smooth, $2(n-k)$-dimensional submanifold of $M$.
\end{thm}

\prf
Let $\lbrace Y_i\rbrace_{i=1}^{2k}$ denote a smooth, local family of
spanning vector fields for $V$ over an open neighborhood $U\subset M$.
Since $V$ is symplectic, the fact that $X_H^V$ is a section of $V$
implies that $\omega(Y_i,X_H^V)$ cannot be identically zero for all $i$
and everywhere in $U$.
Due to the $\omega$-orthogonality of $\PV$, and $\PV Y_i=Y_i$,
$$\omega(Y_i,X_H^V)\;=\;\omega(\PV (Y_i),X_H)\;=\;\omega(Y_i,X_H)
    \;=\;Y_i(H)\;.$$
Thus, with $\underline{F}:=(Y_1(H),\dots,Y_{2k}(H))\in
   C^\infty(U, \R^{2k})$,
it is clear that $\crit\cap U=\underline{F}^{-1}(\underline{0})$.
Since $\underline{F}$ is smooth, Sard's theorem implies
that regular values, having smooth, $2(n-k)$-dimensional
submanifolds of $U$ as preimages,
are dense in $\underline{F}(U)$ \cite{Mi2}. \qed

Next, let us pick a local spanning
family  $\left\lbrace Y_i\in\Gamma(V)\right\rbrace_{i=1}^{2k}$
for $V$ that  satisfies
$$\omega(Y_i,Y_j)\;=\;\tJ_{ij}\;,$$
with
$\tJ:=\left[\begin{array}{cc}0&\1_k\\-\1_k&0\end{array}\right]$.
This choice is always possible.

Furthermore, introducing the
associated family of 1-forms $\lbrace\theta_i\rbrace$
by $\theta_i(\cdot):=\omega(Y_i,\cdot)$,
theorem {~\ref{Pconstr}}
implies that
$$\PV\;=\;\tJ^{ij}Y_i\otimes \theta_j\;,$$
where $\tJ^{ij}$ are the components of $\tJ^{-1}=-\tJ$.
Expanding $X_H^V$ with respect to the basis $\lbrace Y_i\rbrace$ gives
\eqn\label{XHVexp} X_H^V\;=\;\PV(X_H^V)\;=\;-\;Y_i(H)\;\tJ^{ij}\; Y_j\;,\eeqn
where one uses the relationship $\theta_j(X_H^V)=Y_j(H)$
obtained in the proof of theorem {~\ref{Sardthm}}.

\begin{prp}\label{invert}
Under the genericity assumption of theorem {~\ref{Sardthm}},
the $2k\times 2k$-matrix
$[Y_k(Y_i(H))(x_0)]$ is invertible for all  $x_0\in\crit$,
and every local spanning family $\lbrace Y_i\in\Gamma(V)\rbrace$ of $V$.
\end{prp}

\prf
Let us pick a local
basis $\lbrace Y_i\rbrace_1^{2k}$ for $V$, and
$\lbrace Z_j\rbrace_1^{2(n-k)}$ for $V^\perp$, which together span $TM$.
Let $x_0\in\cg$, and assume the generic situation of theorem
{~\ref{Sardthm}}. Because $\cg$ is defined as the set of
zeros of the vector field (~\ref{XHVexp}), the kernel of the linear map
$$\left. dF_i(\cdot)\tJ^{ik}Y_k\right|_{x_0}\;:\;
       T_{x_0}M\;\rightarrow\;V_{x_0}\;,$$
where $F_i:= Y_i(H)$, is precisely $T_{x_0}\cg$,
and has a dimension $2(n-k)$.

In the basis $\lbrace Y_1|_{x_0},\dots,Y_{2k}|_{x_0},Z_1|_{x_0},\dots,
Z_{2(n-k)}|_{x_0}\rbrace$, its  matrix is given by
$$A\;=\;\left[ A_V\;A_{V^\perp}\right]\;,$$
where $A_V:=[Y_i(F_j)\tJ^{jk}Y_k|_{x_0}]$, and
$A_{V^\perp}:=[Z_i(F_j)\tJ^{jk}Y_k|_{x_0}]$.
Bringing $A$ into upper triangular form, $A_V$ is also transformed into upper
triangular form. Because the rank of $A$ is $2k$, and $A_V$ is a
$2k\times 2k$-matrix, its upper triangular form has $2k$
nonzero diagonal elements. Consequently, $A_V$ is invertible,
and due to the invertibility of $\tJ$, one arrives at the
assertion. \qed

\begin{cor}
Let $\cg$ satisfy the genericity assumption of theorem {~\ref{Sardthm}}.
If $V$ is integrable, the intersection of any integral manifold
of $V$ with $\cg$ is a discrete set.
\end{cor}

\prf
The previous proposition implies that generically, integral manifolds of
$V$ intersect $\cg$ transversely.
Their dimensions are
complementary, hence the intersection set is zero-dimensional. \qed

\subsection{Topology of the critical manifold}

\noindent{\bf The Hessian of the constrained system.}
The usual coordinate invariant definition of the Hessian of $H$
corresponds to $\nabla dH$, evaluated at
its critical points
\cite{Jo}, where $\nabla$ denotes the Levi-Civita connection
of the K\"ahler metric $g$.

A generalized Hessian for the
constrained Hamiltonian system can easily be defined along these lines.
Let us write
$\PV^\dagger$ for the dual projector associated to $\PV$, which acts 
on sections of the
cotangent bundle $T^*M$.
Let $\theta$ denote any one-form. Then of course,
$\langle \PV^\dagger\theta, X\rangle =\langle\theta,\PV X\rangle$.
In any local
chart, the matrix of $\PV^\dagger$ is the transpose of the matrix of $\PV$.

The obvious
generalization of the Hessian is the tensor
$\nabla (\PV^\dagger dH)$. It acts as a bilinear form on pairs of vector fields
in terms of
\eqnn\nabla(\PV^\dagger dH)(X,Y)&\equiv&\langle\nabla_X (\PV^\dagger dH),
         Y\rangle \\
         &=&(((\PV)^j_r H_{,j})_{,s} -\Gamma^s_{ri}(\PV)_s^j H_{,j} ) 
         X^r Y^s ,\eeqnn
where $\Gamma^s_{ri}$ are the Christoffel symbols.
The second term in the bracket on the lower line is zero on $\crit$,
because $(\PV)_s^j H_{,j}=0$  on
$\crit$.
The non-vanishing term in $\nabla(\PV^\dagger dH)$ on $\crit$ is determined by the
matrix
\eqn\label{hessgen}K_{rs}\equiv ((\PV)^j_r H_{,j})_{,s} .\eeqn
A straightforward calculation 
shows that
$(\PV)^j_i  K_{jk} = K_{ik}$
holds everywhere on $\crit$.
Obviously, the rank of $K$ is bounded from
above by the rank of $\PV$, corresponding
to the rank $2k$ of $V$.

The corank of
$K|_a$ equals the dimension of the connectivity component of $\crit$
which contains $a$.
This is because for any
$a'\in\crit$ close to $a$, Taylor's theorem implies that
$$((\PV)_s^j H_{,j})(a')=((\PV)_s^j H_{,j})(a)+ K|_a(a'-a) + O(\|a'-a\|^2) .$$
The left hand side and the first term on the right hand
are both zero, since $\crit$ is so defined,
thus $\frac{K(a'-a)}{\|a'-a\|}\rightarrow0$ in the limit
$\|a'-a\|\rightarrow 0$.
That is, the tangent space $T_a\crit$ is equal to  ${\rm ker}K|_a$,
hence the
corank of $K|_a$ is indeed the dimension of the component $\crit_i$.
\\

\subsubsection{An auxiliary gradient-like dynamical
system}\label{auxgradlkflowsssec}

The physical flow of the constrained Hamiltonian system
turns out to of very limited use for the study of the global
topology of $\crit$. This is because invariant sets of the
physical flow not only contain fixed points, but also
periodic orbits.

However, the
auxiliary dynamical system on $M$ defined by
\eqn\label{gradlikeODE}\partial_t\gamma(t)
    =-\left. (\PV\grad H)\right(\gamma(t))\;, \eeqn
for $\gamma:I\subset\R\rightarrow M$,
is an extremely powerful tool for this purpose.
Let us denote its flow by
$$\phi_t\;\in\;{\rm Diff}(M)\;.$$
Its orbits are of course tangent to $V$,
and it is clear that $\Phi_t^c$ and $\phi_t$
share the same critical set $\crit$.

\begin{dfi}
A flow is gradient-like if there exists
a function $f:M\rightarrow\R$ that decreases strictly along all of its
non-constant orbits.
\end{dfi}

\begin{prp}
The flow $\phi_t$ is gradient-like.
\end{prp}

\prf
One can easily check that $H$ decreases strictly along the
non-constant orbits of $\phi_t$,
\eqnn\partial_t H(\gamma(t))&=&\left\langle\; dH(\gamma(t))\;,\;
        \partial_t \gamma(t)\;\right\rangle\\
        &=&-\;g\left(\grad H\;,\;\PV\grad H\right)(\gamma(t))\\
        &=&-\;g\left(\PV\grad H\;,\;\PV\grad H\right)(\gamma(t))\;
        \nonumber\\
        &\leq& 0 . \eeqnn
We have here used the fact that $(g,J,\omega,\PV)$ is a compatible quadruple.
This is the main motivation for the introduction of these quantities.
\qed

It is immediately clear that
$\phi_t$ generates no periodic trajectories,
hence $\crit$ comprises all invariant sets of $\phi_t$.
\\

\noindent{\bf The gradient-like flow close to $\crit$.}
Let us briefly consider the orbits of $\phi_t$ in a tubular
$\epsilon$-neighborhood of $\crit$ (with respect to the
Riemannian distance induced by the K\"ahler metric $g$).
We pick an arbitrary element $a\in\crit$ and
local coordinates $x^i$, with the origin at $a$.
The equations of motion (~\ref{gradlikeODE}) are given by
\eqn\partial_t x^r(t) \;=\;-\;K_{s}^r(a)x^s(t)\;+\;O(\|x\|^2) ,\eeqn
where $K_{js}$ is the 'constrained Hessian' at $a$,
and $K^r_s=g^{rj}K_{js}$, and $\|\cdot\|$ is with respect to
$g$ at $a$.

$P_a$ will from now on denote the matrix of $\PV(a)$. Furthermore,
let $\IPa:V_a\hookrightarrow T_aM$ denote the embedding.
Moreover, we will write
$\K\equiv [K_{k}^j(a)]$ for the Jacobian matrix of $\PV\grad H$ at $a$
in the present chart. From the $g$-orthogonality of $\PV$ follows
straightforwardly that
$P_a A_a=A_a$.

The linearized equations of motion at $a$ read
$\partial_t x(t)=-\K x(t)$,
and satisfy the linearized constraints
$\partial_t x(t)=P_a\partial_t x(t)$.

Let $\bar{\pop}$ denote the $g$-orthoprojector onto ${\rm ker}\K=T_a\crit$,
and let $\IbQa:T_a\crit\hookrightarrow T_aM$ be the embedding.
Its complement $\pop_a=\1-\bar{\pop}_a$ projects $T_a M$ orthogonally onto the fibre
$N_a\crit$ of the normal bundle of $\crit$, and likewise, let
$\IQa:T_aN\hookrightarrow T_aM$ be the embedding. The orbits of the linearized
system are then given by
\eqn x(t)&=& \exp(-t \IPa \K \IQa\pop_a)x_0\\
        &=& \exp(-t \IPa \K\IQa \pop_a)\IQa\pop_a x_0 +
        \IbQa\bar{\pop}_a x_0, \eeqn
where the initial condition is determined by $x(0)=x_0$.
The vector $\bar{\pop}_a x_0\in T_a\crit$ approximately connects $a$
with some critical point close to $a$, while $\pop_a x_0$ lies in $N_a\crit$.
\\

%%\begin{figure}[h]
%%\begin{center}
%%\includegraphics{Fig1.eps}
%%\end{center}
%%\caption{The gradient-like flow in the vicinity of $\cg$. The axes
%%denote the images of the respective projectors, and the dotted lines
%%schematically account for the orbits of the linearized flow (assuming that
%%$\cg$ is an attractor).}
%%\label{Fig1}
%%\end{figure}

\subsubsection{\nit Morse functions and non-degenerate critical manifolds}

Let us recall some standard definitions from
Morse- and Morse-Bott theory that will be needed in the subsequent discussion.

\begin{dfi}
The dimensions of the  zero and negative eigenspaces of the Hessian  of
$f$ at a critical point $a$ are
called the nullity and the index of the critical point $a$.
If all critical points of $f:M\rightarrow\R$ have a zero nullity,
$f$ is called a Morse function, and the index is
then called the Morse index of $a$.

If the critical points of $f$ are not isolated, but elements of
critical manifolds that are non-degenerate in the sense of Bott,
$H$ is called a Morse-Bott function \cite{Bo}.
\end{dfi}

Throughout this section, we will
assume that $H$ is a Morse function.

\begin{dfi}
A connectivity component $\crit_i$ is locally
normal hyperbolic at the point $a\in\crit$
with respect to $\phi_t$ if it is a manifold at $a$, and
if the restriction of $\K$ to the normal space $N_a\crit$ is non-degenerate.
A connectivity component $\crit_i$ is called non-degenerate
if it is a manifold that is everywhere normal hyperbolic with respect
to $\phi_t$.
The index of a non-degenerate
connectivity component $\crit_i$ is
the number of eigenvalues of the constrained Hessian
$\K$ on $\crit_i$ that are contained in the negative half plane.
\end{dfi}

\begin{prp}
In the generic case, the critical manifold is normal hyperbolic with
respect to the gradient-like flow.
\end{prp}

\prf This follows straightforwardly from lemma {~\ref{invert}}. \qed

Let $\crit_i$, $i=1,..,l$ denote the connectivity
components of $\crit=\cup \crit_i$, and let $j_i:\crit_i\rightarrow M$ denote
the embedding.

\begin{prp}
Assume that $\crit$ satisfies the genericity assumption in the sense of
Sard's theorem. Then, if $H:M\rightarrow\R$ is a Morse function,
$H_i:=H\circ j_i:\crit_i\rightarrow\R$ is also a Morse function.
A point $x\in \crit_i$ is a critical point
of $H_i$ if and only if it is a critical point of $H$.
\end{prp}

\prf
The fact that every critical point of $H$
is a zero locus of $\PV\grad H|_a$,
and thus an element of $\crit$, is trivial.

To prove the converse statement,
let $\crit_i$ be generic, so it has a dimension $2(n-k)$.
Normal hyperbolicity implies that the restriction of
$P_a\IQa\pop_a \IPa P_a$ to
the normal space $N_a\crit_i$ defines an invertible
map for all $a\in\crit_i$.
Let $a$ now denote an extremum of $H|{\crit_i}$.
Then,
$$\langle dH,v\rangle|_a = g(\grad H,v)|_a=0 $$
holds for all $v\in T_a\crit_i$.
Consequently,
$\grad H|_a$ is a vector in $N_a\crit$, and thus,
$\grad H|_a=\pop_a\grad H|_a$.
Moreover, by definition of $\crit$,
the condition $P_a\grad H|_a = 0$ is satisfied,
which in turn implies
$\pop_a\IPa P_a\IQa\pop_a\grad H|_a =0$. Hence, since $\pop_a
\IPa P_a\IQa\pop_a:N_a\crit_i\rightarrow N_a\crit_i$ is invertible,
it follows that
$\grad H|_a=0$.

The Hessian of the restriction of $H$ at any critical point of $H_i$
is always nondegenerate, thus $H_i$ is by itself
a Morse function on $\crit_i$. \qed

\begin{cor}
The critical points of $H|_{\cg}$ are precisely the critical points
of $H:M\rightarrow\R$. Let $\crit_i$ be a
non-generic
connectivity components with manifold structure
that is normal hyperbolic. Then,
$\crit_i\subset\Sigma_{H(\crit_i)}$.
\end{cor}

\prf The first assertion follows trivially from the previous
proposition.

Next, let us assume that $\crit_i$ is some non-generic connectivity
component of $\crit$ with manifold structure that is normal hyperbolic
(which is in no sense
necessitated by the system). Then, the proof of the previous
proposition can be applied to the situation for $\crit_i$, and demonstrates
that there are no extrema of $H|_{\crit_i}$. Thus,
$\crit_i$ is a submanifold of the level surface
$\Sigma_{H(\crit_i)}$. \qed

\subsubsection{\nit Index pairs and relative (co-)homology}

Let us next focus on the global topology
of $\crit$, for assumptions that are weaker than genericity.
In fact, we will only assume that the generic connectivity
components $\cg\subset\crit$ are non-degenerate,
compact and without boundary, and that $\crit\setminus\cg$ is compact.
In this situation, a generalization of the Morse-Bott inequalities can be
derived for $\crit$, based on an application of the theory of
C. Conley and E. Zehnder for flows on Banach spaces with compact invariant
sets.
We will only give a very short survey of Conley-Zehnder (CZ) theory,
and refer the reader to the original article \cite{CoZe}.

\begin{dfi}
Let $\crit_i$ be any compact component of $\crit$.
An index pair associated to $\crit_i$
is a pair of compact sets
$(N_i,\tilde{N}_i)$ that possesses the following properties.
The interior of $N_i$  contains $\crit_i$, and moreover,
$\crit_i$ is the maximal invariant
set under $\phi_t$ in the interior of $N_i$.
$\tilde{N}_i$ is a compact subset of $N_i$ that has empty intersection with
$\crit_i$,
and the trajectories of all points in $N_i$ that leave
$N_i$ at some time under the gradient-like flow $\phi_t$ intersect
$\tilde{N}_i$.
$\tilde{N}_i$ is called the exit set of $N_i$.
\end{dfi}

The $p$-th  relative homology group $H_p(X,A)$ of a pair of manifolds
$A\subset X$ is defined as follows. Let $C_p(X)$ be the group
of $p$-cycles of $X$. Since $C_p(A)$ is a subgroup of $C_p(X)$,
the quotient group $C_p(X)/C_p(A)$ is well-defined. Let  $B_p(X,A)$
denote the subgroup of $C_p(X)/C_p(A)$ which consists of boundaries. The
associated quotient group is the $p$-th relative homology group
$H_p(X,A)$.

The $p$-th  relative de Rham cohomology group
$H^p(X,A)$ consists of cohomology classes that are represented by closed
$p$-forms on
$X$ whose restriction to $A$ (via the pullback of the embedding $A\rightarrow
X$) is exact. If $X$ and $A$ are both orientable, the coefficients of the
relative homology and cohomology groups can be picked from $\R$ or ${\bf Z}$, and
otherwise from ${\bf Z}_2$.
\\

\noindent{\bf The relative cohomology of an index pair.}
It was proved in \cite{CoZe} that the homotopy type of the
pointed space
$N_i/ \tilde{N}_i$ only depends on $\crit_i$, so that
$H^*(N_i,\tilde{N}_i)$ is independent of the particular choice of index pairs
(the space $N_i/ \tilde{N}_i$ is obtained from collapsing the subspace
$\tilde{N}_i$ of $N_i$ to a point). The equivalence class
$[N_i/ \tilde{N}_i]$ of pointed topological spaces under homotopy
only depends on
$\crit_i$, and is called the {\em Conley index} of $\crit_i$.
In the present analysis, we will only consider flows exhibiting normal
hyperbolic critical manifolds. In this special case, the result essentially
reduces to a generalization of Morse-Bott theory.
We will now determine the relative cohomology
$H^*(N_i,\tilde{N}_i)$ of an index pair $(N_i,\tilde{N}_i)$
associated to a generic connectivity component $\crit_i$.

\begin{prp}
Let $\crit_i\subset\cg$ be a generic connectivity component that
is compact and without boundary, and let
$(N_i,\tilde{N}_i)$ denote any associated index pair. Then,
\eqn H^{q+\mu_i}(N_i,\tilde{N}_i)\;\cong\; H^q(\crit_i)\;,\eeqn
where $\mu_i$ is the index of $\crit_i$, and $q=0,\dots,dim(\crit_i)$.
\end{prp}

\prf
Let us consider, for some sufficiently small $\epsilon_0>0$,
a compact tubular
$\epsilon_0$-neighborhood $U$ of $\crit_i$ (of dimension $2n$), and let
$$W_U^{cu}(\crit_i)\;:=\;(W^-(\crit_i)\cup \crit_i)\cap U$$
denote the intersection of the center unstable manifold of $\crit_i$
with $U$. $W^-(\crit_i)$ denotes the unstable manifold of $\crit_i$.
Pick some small, positive $\epsilon <\epsilon_0$, and let $U_\epsilon$
be the compact tubular $\epsilon$-neighborhood
of $W_U^{cu}(\crit_i)$ in $U$.

It is clear that letting $\epsilon$ continuously go to zero,
a homotopy equivalence of tubular neighborhoods is obtained,
for which $W_U^{cu}(\crit_i)$ is a deformation retract.
Let
$$U_\epsilon^{out}\;:=\;\partial U_\epsilon \cap \phi_\R(U_\epsilon)$$
denote the intersection of $\partial U_\epsilon$ with all orbits of the
gradient-like flow that contain points in $U_\epsilon$.
Then, evidently, $(U_\epsilon,U_\epsilon^{out})$ is an index pair for
$\crit_i$, and by letting $\epsilon$ continuously go to zero,
$U_\epsilon^{out}$ is homotopically retracted to
$\partial W_U^{cu}(\crit_i)$.

Thus, homotopy invariance implies that the relative cohomology groups
obey
\eqnn H^*(U_\epsilon,U^{out}_\epsilon)\cong
      H^*(W_U^{cu}(\crit_i),\partial W_U^{cu}(\crit_i)).\eeqnn
Due to the normal hyperbolicity of $\crit_i$ with respect to the
gradient-like flow, $W_U^{cu}(\crit_i)$ has a constant dimension
$n_i+\mu(\crit_i)$ everywhere, where $n_i={\rm dim}\crit_i$.
Therefore, one obtains from Lefschetz duality \cite{DuFoNo} that
$$H^{n_i+\mu_i-p}(W_U^{cu}(\crit_i),\partial W_U^{cu}(\crit_i))\cong
        H_p(W_U^{cu}(\crit_i)\setminus\partial W_U^{cu}(\crit_i)) ,$$
where $\mu_i=\mu(\crit_i)$, the index of $\crit_i$.
It is clear that $\crit_i$ is a deformation retract of the interior of
$W_U^{cu}(\crit_i)$, so that the
respective cohomology groups are isomorphic.

Due to $dim(\crit_i)=n_i$, Poincar\'e duality implies
$$H_p(W_U^{cu}(\crit_i)\setminus\partial W_U^{cu}(\crit_i))
      \cong H_p(\crit_i)\cong H^{n_i-p}(\crit_i),$$
so that with $q:=n_i-p$,
\eqn\label{holy}H^{q+\mu_i}(U_\epsilon,U^{out}_\epsilon)
    \cong H^q(\crit_i) ,\eeqn
which proves the claim.
\qed

\subsubsection{\nit The Conley-Zehnder inequalities}

Here follows a very rough summary of the main steps in the
derivation of the Conley-Zehnder inequalities, \cite{CoZe}.

\begin{dfi} Let $I$ denote a compact invariant invariant set under
$\phi_t$.
A  Morse decomposition of
$I$ is a finite, disjoint family of compact, invariant subsets
$\lbrace M_1,\dots,M_n\rbrace$ that satisfies the following requirement on
the
ordering. For every $x\in I\setminus \cup_i
M_i$, there exists a pair of indices $i<j$, such that
$$\lim_{t\rightarrow-\infty}\phi_t(x)\subset M_i \hspace{2cm}
    \lim_{t\rightarrow \infty}\phi_t(x)\subset M_j.$$
Such an ordering, if it exists, is called admissible, and the $M_i$ are
called Morse sets of $I$.
\end{dfi}

A central result proved in \cite{CoZe} is that for every compact
invariant
set $I$ admitting an admissibly ordered Morse decomposition, there exists
an increasing sequence of compact sets $N_i$ with $N_0\subset
N_1\subset\dots\subset N_m$, such that $(N_i,N_{i-1})$ is an index pair for
$M_i$, and $(N_m,N_0)$ is an index pair for $I$.

Consider compact manifolds $A\supset B \supset C$. The exact sequence of
relative cohomologies
$$\dots\stackrel{\delta^{k-1}}{\rightarrow} H^k(A,B)\rightarrow
       H^k(A,C)\rightarrow H^k(B,C)      \stackrel{\delta^k}{\rightarrow}
       H^{k+1}(A,B)\rightarrow\dots$$
implies, in a standard fashion known from
Morse theory, that, with $r_{i,p}$ denoting the rank of
$H^p(N_i,N_{i-1})$, the identity
$$\sum_{i,p}\lambda^p r_{i,p}=\sum_p B_p \lambda^p + (1+\lambda)\Q(\lambda) $$
holds for the indicated Poincar\'e polynomials \cite{Jo}.
Here, $B_j$ is the $j$-th Betti number of the index pair
$(N_m,N_0)$ of $I$, and
$\Q(\lambda)$ is a polynomial in $\lambda$
with non-negative integer coefficients.
These are the strong Conley-Zehnder inequalities; a corollary that
straightforwardly follows, due to the positivity of the coefficients of
$\Q(\lambda)$,
is that
$$\sum_i r_{i,p}\geq B_p$$
holds. These are the weak Conley-Zehnder inequalities.

If the symplectic manifold $M$ is  compact and  closed, and if
$\crit$
is assumed to be non-degenerate, the following results
arise from application of the CZ inequalities. The invariant set $I$ can be
chosen to be equal to $M$. So we let
$N_m=M$ and $N_0=\emptyset$ denote the top and bottom elements of the
sequence defined above. Furthermore, we order the connected elements of
$\crit$
according to the descending values of the maximum of $H$ attained on each
$\crit_i$. Then it follows that $\crit$ furnishes a Morse decomposition for
$M$. The homology groups of $M$ are isomorphic to the relative homology
groups
of the index pair $(N_m,N_0)$. So the numbers
$B_p$ are the Betti numbers of the compact symplectic manifold $M$.

The number $r_{i,p}$ is the rank of the $p$-th relative cohomology group of
the
non-degenerate critical manifold $\crit_i$, the index $i$ being determined by
the Morse decomposition. From (~\ref{holy}), we deduce that
$$r_{i,p} = {\rm dim} H^{i,p-\mu_i}(\crit_i) $$
(we recall that $\mu_i$ is the index of $\crit_i$).
In other words,
$r_{i,p}$ is the $(p-\mu_i)$-th Betti number of $\crit_i$, written
$B_{i,p-\mu_i}$ in brief. Assuming that the number of connected components of
$\crit$ is finite, the Conley-Zehnder inequalities thus result in
\eqn\label{CZ0}\sum_{i,p} B_{i,p}\;\lambda^{p+\mu_i}\;=\;
     \sum_p \;B_p\; \lambda^p +
     (1+\lambda)\;\Q(\lambda) ,\eeqn
which implies that
\eqn\label{weakCZ}\sum_{i} B_{i;p-\mu_i} \geq B_{p}\eeqn
is satisfied.  So the
global topology of $M$ enforces lower bounds on the ranks of the homology
groups
of the connected components of $\crit$. Setting the variable $\lambda$ equal
to $-1$, one obtains that
$$\sum_{i,p} (-1)^{p+\mu_i} B_{i,p} = \sum_{i} (-1)^{\mu_i} \chi(\crit_i)
        = \chi(M) ,$$
where $\chi$ denotes the Euler characteristic.
\\

\noindent{\bf Remark.} In the case of mechanical systems, the phase space
of the relevant constrained Hamiltonian system is
non-compact,
and the critical manifold is generally unbounded.
Therefore, the arguments used here do not apply.
However, since in that case,
$M$ and $\crit$ are vector bundles, we are nevertheless able
to prove results that are fully analogous to (~\ref{CZ0}).
\\

\noindent{\bf Generic connectivity components.}
One can actually prove a stronger result than (~\ref{CZ0}),
given the special structure of the system at hand.

\begin{prp}\label{CZineqprp1nongeneric}
Assume that $\crit\setminus\cg$ is a disjoint union of $C^1$-manifolds. Then,
\eqn\label{CZ1}\sum_{\stackrel{i,p}{\crit_i\in\cg}}
       B_{i,p}\;\lambda^{p+\mu_i}\;=\;\sum_p \;B_p\;
      \lambda^p \;+\; (1+\lambda)\;\tilde{\Q}(\lambda) \;. \eeqn
$B_{i,p}$ are the $p$-th
Betti numbers of the connectivity components $\crit_i$ of
$\cg$, $B_p$ are the Betti numbers of $M$,
and $\tilde{\Q}$ is a polynomial with non-negative integer coefficients.
\end{prp}

\prf
We will show that $X_H^V$ can be suitably deformed such that
the components of $\crit\setminus\cg$ can be smoothed away.
To this end,
we construct an auxiliary, continuous vector
field $X_\epsilon$, corresponding to a small deformation of the
gradient-like vector field
$\PV\grad H$. We pick a small positive, real number
$\epsilon$, and consider the compact neighborhoods
\eqn\label{neighb}U_\epsilon(\crit_i)=\lbrace x\in M\mid
      dist_g(x,\crit_i)\leq\epsilon\rbrace\eeqn
of connectivity components $\crit_i\subset \crit\setminus\cg$.
Here, $dist_g$ denotes the
Riemannian distance function
induced by the K\"ahler metric $g$.
The vector field
$X_\epsilon$ shall be given by $\PV\grad H$ in
$M\setminus U_\epsilon(\crit_i)$, and
in the interior of every $U_\epsilon(\crit_i)$ with
$\crit_i\subset\crit\setminus\cg$,
\eqn\label{kiwi}X_\epsilon|_x=\PV\grad H|_x +\epsilon h(x)\grad
   H|_x\;.\eeqn
Here, we have introduced a $C^1$-function
$h:U_\epsilon(\crit_i)\rightarrow[0,1]$ obeying
$h|_{\crit_i}=1$ and $h|_{\partial U_\epsilon(\crit_i)}=0$,
that is strictly monotonic along the
the flow lines generated by $\PV\grad H$. Such a function
$h$ exists because $\crit\setminus\cg$ is a disjoint union
of $C^1$-manifolds.

We have proved above that for all $\crit_i\subset\crit\setminus\cg$,
$\grad H$ is strictly non-zero in $U_\epsilon(\crit_i)$.
Now let us consider
$$g(X_\epsilon,\grad H)=(\|\PV\grad H\|_g^2)(x)+\epsilon h(x) (\|\grad
      H\|_g^2)(x) .$$
Here we have used the $g$-symmetry of $P$, and
$\|X\|_g^2\equiv g(X,X)$.
The first term on the right hand side is non-zero on
the boundary of $U_\epsilon(\crit_i)$, while the second term vanishes.
Moreover, the second term is non-zero everywhere in the interior of
$U_\epsilon(\crit_i)$. Therefore, $X_\epsilon$ vanishes nowhere in
$U_\epsilon(\crit_i)$.

This implies that the vector field $X_\epsilon$ is a deformation of
$\PV\grad H$ that only exhibits $\cg$ as a critical set.
Notice that the generic components cannot be removed in this manner,
since they contain critical points of $H$.

The scalar product of $X_\epsilon$ with $\grad H$ is not only non-zero,
but also strictly positive in every
$U_\epsilon(\crit_i)$, which shows that $X_\epsilon$ is a gradient-like
flow (the left hand side of the above expression corresponds to the time
derivative of $H$ along orbits of
$X_\epsilon$ in $U_\epsilon(\crit_i)$).

Clearly, the order of magnitude of $\|X_\epsilon-\PV\grad H\|_g$
is at most $O(\epsilon)$ everywhere on $M$.
Consequently, it is possible to pick $X_\epsilon$
arbitrarily close to $\PV\grad H$ in the $\|\cdot\|_\infty$-norm on
$\Gamma(TM)$ that is induced by $\|\cdot\|_g$.

Carrying out
the Conley-Zehnder construction with respect to the flow generated by
$X_\epsilon$ yields (~\ref{CZ1}).
This result does not require the assumption of normal hyperbolicity
on $\crit$. \qed

\subsubsection{\nbf Proof of (~\ref{CZ1}) using the Morse-Witten complex}

We will now give a different proof of (~\ref{CZ1}) based on
the existence of the Morse-Witten complex.
Our motivation is, on the one hand, to clarify the orbit structure
of the gradient-like system, and, on the other
hand, to further comment on the "position" of $\cg$ in $M$ relative to
the equilibria of the free system (corresponding to the critical
points of $H$).

To this end, let us first make some brief
general remarks about Morse theory, cf. \cite{Bo1}.

In the basic setting, there is a Morse function
$f:M\rightarrow\R$ on a compact manifold $M$, without boundary, say.
The index of an
isolated critical point precisely is its Morse index,
hence the strong Conley-Zehnder
inequalities (~\ref{CZ1}) yield
$$\sum_p \lambda^p N_p =\sum_p \lambda^p B_p + (1+\lambda)\Q(\lambda) ,$$
where $N_p$ is the number of critical points of $f$ with a Morse index $p$.
$B_p$ is the $p$-th Betti number of $M$, and $\Q(t)$ is a polynomial with
non-negative integer coefficients.
These are the strong Morse inequalities,
from which the weak Morse inequalities $N_p\geq B_p$ are immediately inferred.
The standard proof uses the gradient flow generated by $-\grad f$ with respect
to an arbitrary auxiliary metric.

If $f$ is not assumed to be a Morse, but more generally, a Morse-Bott
function  \cite{Bo},
the gradient flow of the vector field $-\grad f$ can again be used,
but then
to derive the strong Conley-Zehnder
inequalities for the critical
manifolds of $f$.

Morse theory, which has been a classical topic in differential topology since
the 1930's, experienced a tremendous new increase of 
activity in the early
1980's, due to the seminal work of  E. Witten
\cite{Wi}. His proof of the weak Morse inequalities is based on a deformation
and localization argument in Hodge theory, interpreted as supersymmetric
quantum
mechanics. Subsequently, the proof of the Morse-Bott inequalities
using heat kernel
methods has been given by J.-M. Bismut in \cite{Bi}.

This new approach to the strong Morse inequalities is based on the
differential complex constructed from the the set of
critical points of $f$, which is
often referred to as the Morse-Witten complex.
At least Milnor, Smale and Thom
independently  have arrived at some form of it  already earlier
\cite{Bo1,Fl,Sw}. The (co-)homology of the Witten complex is isomorphic to
the
singular (de Rham co-)homology of
$M$, and straightforwardly implies the strong Morse inequalities.
A proof of the Morse and Morse-Bott inequalities that is based on the
construction of the Morse-Witten complex is given in \cite{AuBr}.

The Morse-Witten complex has been generalized to infinite dimensional 
systems by
A. Floer \cite{Fl}, which has led to extremely fruitful applications.
For instance, the celebrated Arnol'd conjecture has been proved by use of
Floer homology, \cite{HoZe}. A beautiful survey is given in \cite{Ze}.

We will now discuss an alternative proof of (~\ref{CZ1})
that is based on the Morse-Witten complex for {\em non}-degenerate  Morse
functions.
\\

\noindent{\bf The Morse-Witten complex.}
Let $M$ be a compact, closed, orientable and smooth manifold,
together with a Morse function $f:M\rightarrow\R$.
We let $\C^p$ denote the free $\Z$-module generated
by the critical points of $f$ with a Morse index $p$.
The set $\C=\oplus_p \C^p$ is the free $\Z$-module generated by the
critical points of $f$, which is graded by their Morse indices.
There exists a natural coboundary operator
$$\delta:\C^p\rightarrow\C^{p+1}$$
that obeys $\delta^2=0$, which we will define below.

\begin{thm}\label{CohWitdRhthm}
The cohomology of the differential complex $(\C,\delta)$
is isomorphic to the de Rham
cohomology of $M$,
${\rm ker}\delta / {\rm im}\delta \cong H^*(M,\Z)$.
\end{thm}

A proof of this theorem has been given by Floer in \cite{Fl}
based on Conley-Zehnder theory.
Other proofs can be found in \cite{AuBr,Sw}.
The argument in the original publication \cite{Wi}
is based on the quantum mechanical tunneling effect.

The coboundary operator is defined as follows
\cite{AuBr,Bo1,Fl,Wi}.
We denote the unstable and
stable manifold of a critical point $a$ of $f$ under the
gradient flow by $W_a^-$ and $W_a^+$, respectively, and assign an
arbitrary orientation to every $W_a^-$.
Since $M$ is assumed to be oriented,
the orientation of $W_a^-$
at every critical point $a$ induces an orientation of $W_a^+$.
The set of
Morse functions for which the stable and unstable manifolds
intersect
transversely is dense in $C^\infty(M)$.
Thus, we may
generically assume that all $W_a^-$ and $W_{a'}^+$ intersect
transversely.
The dimension of $W_a^-$ equals the Morse index $\mu(a)$ of $a$,
and the dimension of the intersection
$$M(a,a')\equiv W_a^-\cap W_{a'}^+$$
is given by  ${\rm max}\lbrace\mu(a)-\mu(a'),0\rbrace$.

In order to define the coboundary operator,
let us consider pairs of critical points
$a$ and
$a'$, whose relative Morse index has the value $1$, say
$\mu(a)=p+1$ and $\mu(a')=p$.
It immediately follows that $M(a,a')$ is a
finite collection of gradient lines that connect $a$ with $a'$.

The intersection of $M(a,a')$ with every regular level
surface $\Sigma_c$ of $f$ with $f(\Sigma_c)=c$ lying between
$f(a)$ and $f(a')$ is transverse,
and consists of a  finite collection of isolated points.
The hypersurface $\Sigma_c$,
being a level surface of $f$, is orientable,
\cite{Hi}, so we pick the orientation, which, combined with the section
$\grad f$ of its normal bundle, shall agree with the orientation of $M$.

In addition, the intersection both of
$W_a^-$ and $W_{a'}^+$ with $\Sigma_c$ is
transverse, and the submanifolds
$$W_{a,c}^-\equiv W_a^-\cap\Sigma_c\hspace{2cm}
     W_{a',c}^+\equiv W_{a'}^+\cap\Sigma_c$$
of $\Sigma_c$ are smooth, compact and closed.
In addition, their dimensions add up to the dimension of $\Sigma_c$.
To every point $b$ of the
set $M(a,a')\cap\Sigma_c=W_{a,c}^-\cap W_{a',c}^+$, one assigns the number
$\gamma(b)=1$ if the induced orientation of
$$T_b\Sigma_c = T_b W_{a,c}^-\oplus T_b W_{a',c}^+$$
agrees with the one picked for $\Sigma_c$,
and $\gamma(b)=-1$ otherwise.

\begin{dfi}
The sum
$$\langle a,\delta a'\rangle \equiv \sum_{b\in M(a,a')\cap\Sigma_c}\gamma(b)$$
is the  intersection number
$\sharp(W_{a,c}^-,W_{a',c}^+)$ of the
oriented submanifolds $W_{a,c}^-$ and
$W_{a',c}^+$ of $\Sigma_c$,  \cite{Hi}.
If for a pair $a$ and
$a'$ of critical points with a relative Morse index $1$,
this intersection number is nonzero, they will be said to be
effectively connected (by gradient lines).
\end{dfi}

\begin{dfi}
The coboundary operator of the Morse-Witten complex is the $\Z$-linear map
$\delta:\C^p\rightarrow\C^{p+1}$ defined by
$$\delta a' = \sum_{\mu(a)=p+1} \langle a,\delta a'\rangle a .$$
\end{dfi}

The coboundary operator satisfies $\delta^2 = 0$, and its
cohomology is isomorphic to the de Rham cohomology of $M$.
Proofs of this statement can be found in
\cite{AuBr,Fl,Wi}.
The image of a critical point of Morse index $p$ under the
coboundary map consists of the critical points of Morse index $p+1$ to
which it is effectively connected.
One also refers to the intersection number $\langle a,\delta a'\rangle$
as the {\em matrix element} of the
$\delta$-operator with respect to the elements $a$ and $a'$ of $\C$.

The existence of the Morse-Witten complex straightforwardly implies the strong
Morse inequalities, as one deduces from the following fact.
Let us denote the $p$-th cocyle group by $\Z^p\subset\C^p$,
which is defined as the intersection of  ${\rm ker}\delta$
with $\C^p$, and let $\B^p\subset\C^p$, the $p$-th coboundary group, denote
the image of $\C^{p-1}$ under $\delta$.
Clearly, $\B^p$ is a subset of
$\Z^p$ because $\delta$ is nilpotent, therefore the set
$\H^p=\Z^p\setminus\B^p$ is well-defined. It is the $p$-th
cohomology group of the Morse-Witten complex.
Since by theorem {~\ref{CohWitdRhthm}}, the cohomology of the Morse-Witten
complex is isomorphic to the de Rham cohomology of $M$,
the rank of $\H^p$ coincides with the
$p$-th Betti number $B_p(M)$ of $M$.

The image of $\C^p$ in $\C^{p+1}$ under $\delta$ is given by
$\B^{p+1}$. Denoting the preimage of $\B^{p+1}$ in $\C^p$ by
$\delta^{-1}(\B^{p+1})$, which is isomorphic to $\B^{p+1}$, one has
$$\C^p=\H^p\oplus \B^p\oplus\delta^{-1}(\B^{p+1}), $$
so that
$${\rm dim}\C^p= B_p(M)+ {\rm dim}\B^p + {\rm dim}\B^{p+1} .$$
The dimension of $\C^p$ equals the number $N_p$ of critical points of
$f$ with a Morse index $p$. Multiplying both sides of the equality sign with
$\lambda^p$, and summing over $p$, one finds
\eqn\sum \lambda^p N_p &=& \sum \lambda^p B_p(M)+ \sum \lambda^p({\rm dim}\B^p + {\rm
                    dim}\B^{p+1})\nonumber\\ \label{oregon}
             &=&\sum \lambda^p B_p(M)+ (1+\lambda)\sum \lambda^{p-1}{\rm dim}\B^p  \eeqn
(notice here that both $\B^0$ and $\B^{2n+1}$ are empty).
These are the strong Morse inequalities, and the polynomial ${\cal
Q}(t)$ defined at the beginning of this section now has a very definitive
interpretation:
$${\cal Q}(\lambda)=\sum \lambda^{p-1}{\rm dim}\B^p .$$
Evidently, ${\rm dim}\B^p$ is the number of critical points of Morse index
$p$ that are effectively connected to critical points of Morse index
$p-1$ via gradient lines of $f$.
\\

\noindent{\bf Comparing the Morse-Witten complexes of $(M,H)$ and
$(\cg,H|_{\cg})$.}
Let us next relate the Morse-Witten complexes of $(M,H)$ and
$(\cg,H|_{\cg})$ to each other.
To this end, it is necessary to recall that the generic connectivity components
$\cg$ of $\crit$ contain all critical points of $H$,
but no other conditional extrema, and that they are
necessarily normal hyperbolic with respect to $\phi_t$.

Let $\A_i:=\lbrace a_{i,1},\dots,a_{i,m}\rbrace$ denote the set of
critical points of $H$ that are contained in $\crit_i$, and let
$\mu(a_{i,r})$
be the associated Morse indices of $H:M\rightarrow\R$.
Furthermore, let $H_i\equiv H|_{\crit_i}$ denote the restriction of the
Hamiltonian to $\crit_i$.
We have previously shown that the map $H_i:\crit_i\rightarrow \R$
is a Morse function,
whose critical points are precisely the elements of $\A_i$.
Furthermore, it is clear that
the number of negative eigenvalues of the Hessian of $H$
at any element of $\A_i$,
whose eigenspaces are normal to $\crit_i$,
equals $\mu(\crit_i)$, the index of $\crit_i$.

The Morse index of $a_{i,r}$ with respect to
$H_i$ is thus $\mu(a_{i,r})-\mu(\crit_i)$.
The Morse-Witten
complex associated to $\crit_i$ is defined in terms of the free $\Z$-module
generated by the elements of $\A_i$,
which is graded by the Morse indices $p$ of
the critical points of $H_i$,
$$\C_i=\oplus_p \C^p_i.$$
To define the coboundary operator $\delta_i:\C^p_i\rightarrow\C_i^{p+1}$,
one uses the gradient flow on $\crit_i$ generated by $H_i$.
One then concludes that
\eqn{\rm ker}\delta_i / {\rm im}\delta_i \cong H^*(\crit_i,\Z) .\eeqn
Application of (~\ref{oregon}) shows that for every $\crit_i\in\cg$,
\eqn\label{ontario}\sum_{p} \lambda^p N_{i,p} &=&
         \sum_{p} \lambda^p B_p(\crit_i)+
        (1+\lambda)\sum_{p} \lambda^{p-1}{\rm dim}\B^p_i ,\eeqn
where $\B^p_i$ is the $p$-th coboundary group of the Morse-Witten complex of
$\crit_i$, and $N_{i,p}$ is the number of critical points of $H_i$ on
$\crit_i$ whose Morse index is $p$.

Since every critical point of $H$ lies on precisely one generic
component $\crit_i$, the number $N_q$ of critical points of $H$
with a Morse index $q$ is given by
$$N_p=\sum_{i}N_{i;p-\mu(\crit_i)} .$$
Multiplying both sides of (~\ref{ontario}) with
$\lambda^{\mu(\crit_i)}$, and summing over $i$, one obtains
$$\sum_{i,p;\crit_i\in\cg}\lambda^{\mu(\crit_i)+p}N_{i,p}=
        \sum_{i,p;\crit_i\in\cg}\lambda^{\mu(\crit_i)+p}
         B_p(\crit_i)+(1+\lambda)\sum_{\crit_i\in\cg}
         \lambda^{\mu(\crit_i)+p-1}{\rm
         dim}\B^p_i,$$
which becomes, after reindexing $\mu(\crit_i)+p\rightarrow q$,
$$\sum_q \lambda^q N_q=\sum_{i,q;\crit_i\in\cg}\lambda^q
         B_{q-\mu(\crit_i)}(\crit_i)+(1+\lambda)\sum_{i,q;\crit_i\in\cg}
         \lambda^{q-1}{\rm dim}\B_i^{q-\mu(\crit_i)} .$$
Combining this result with the strong Morse inequalities
$$\sum_q \lambda^q N_q=\sum_q \lambda^q B_q(M)+(1+\lambda)\sum_q
         \lambda^{q-1}{\rm dim}\B^q ,$$
one obtains that
\eqnn\sum_{i,p;\crit_i\in\cg}\lambda^q
         B_{q-\mu(\crit_i)}(\crit_i)=\sum_q \lambda^q B_q(M)
         \hspace{3cm}\\
         +\;(1+\lambda)\sum_q \lambda^{q-1}
         \left({\rm dim}\B^q-\sum_{\crit_i\in\cg}{\rm
         dim}\B_i^{q-\mu(\crit_i)}\right)\;.\eeqnn
One observes that formula (~\ref{CZ1}) implies that
the polynomial which is multiplied by $(1+\lambda)$ has
non-negative integer coefficients.

Conversely, if one can prove that for all $q$,
\eqn\label{wonder}{\rm dim}\B^q \geq \sum_{\crit_i\in\cg}{\rm
          dim}\B_i^{q-\mu(\crit_i)}\eeqn
holds, one would also obtain an alternative proof of (~\ref{CZ1}).
The main observation here is that the left hand side is defined in terms of the operator
$\delta$ associated to $(M,H)$, while the right hand side is defined
in terms of the operators $\delta_i$ associated to all $(\crit_i,H_i)$  with
$\crit_i\in\cg$.

The quantity ${\rm dim}\B_i^{q-\mu(\crit_i)}$ denotes the number of critical
points of $H$ with a Morse index $q$ in $\crit_i$,
which are effectively connected to critical points of Morse
index $p+1$ in $\crit_i$ via
gradient lines of the Morse function $H_i$ on $\crit_i$.
Therefore, the sum on the right hand side of (~\ref{wonder}) equals the
number of those critical points of $H$ with a Morse index $q$, which are
effectively connected to critical points of Morse index $q+1$ via gradient
lines of the functions $H\circ j_i$ on all generic
$\crit_i$; here, $j_i:\crit_i\rightarrow M$ is
the inclusion map.
\\

\noindent{\bf Proof of inequality (~\ref{wonder}).}
We will now prove inequality
(~\ref{wonder}) by relating the coboundary
operators of the Morse-Witten complexes of $(M,H)$ and
$(\cg,H|_{\cg})$ to each other.
Apart from the fact that this will establish a different proof of the
Conley-Zehnder inequalities, this will also largely clarify the
orbit structure of the auxiliary gradient-like system.
\\

\noindent{\bf Proof strategy.} We will construct a particular
homotopy of vector fields
$v_\sig $, with $\sig \in[0,1]$, that generate gradient-like flows.
Their zeros will be independent of
$\sig $, and hyperbolic.
The vector fields interpolated by $v_\sig $ are, for $\sig =1$,
$v_1=\grad H$,
so that the zeros of $v_\sig $ are precisely the critical
points of $H$, and for $\sig =0$,
$v_0$ is a vector field that is tangent to every
$\cg$.
For every $\sig \in[0,1]$, we will construct a coboundary
operator via the one-dimensional integral curves of $v_\sig $ that connect
its zeros.
These coboundary operators are independent of $\sig $, and act on the
free $\Z$-module $\C$ of the Morse-Witten complex associated to $(M,H)$.
The desired estimate (~\ref{wonder})
then follows from a simple dimension argument.
\\

\noindent{\bf Construction of $v_0$.}
We require $v_0$ to be gradient-like,
and tangent to $\cg$.
Furthermore, the zeros of $v_0$ shall be hyperbolic,
and shall coincide with the critical points of $H$.
Consequently, the dimension of any
unstable manifold of the flow generated by $-v_0$ equals the Morse
index of the critical point of $H$ from which it emanates.

To this end, we use the vector field
$X_\epsilon$ constructed in the proof of proposition
{~\ref{CZineqprp1nongeneric}}. We recall that it
has been obtained by suitably deforming $\PV\grad H$
such that all elements $\crit_i$ of
$\crit\setminus\cg$ are removed.
Furthermore, let us introduce compact $\epsilon$-neighborhoods
$U_\epsilon(\cg)$ of the generic connected components of $\crit$
in the way demonstrated for (~\ref{neighb}).

It is now possible to extend the projector
$\bar{Q}:T_{\cg}M\rightarrow T\cg$ that has been introduced in
{~\ref{auxgradlkflowsssec}}
over the whole embedding space $TU_\epsilon(\cg)$.
To this end, we pick an arbitrary smooth distribution
$W$ over the base manifold $U_\epsilon(\cg)$, whose fibres over
$\cg$ shall coincide with the corresponding fibres of $T\cg$.
Admitting a slight abuse of notation,
we let $\bar{Q}$ denote the $g$-orthogonal projector $TM\rightarrow W$.
Clearly, evaluating $\bar{Q}$ in any $a\in\cg$ gives the projector
$\bar{Q}_a:T_aM\rightarrow T_a\cg$ discussed in {~\ref{auxgradlkflowsssec}}.
Because $W$ is smooth,
$\bar{Q}$ and its orthogonal complement
$Q$ are both smooth tensor fields.

We define the vector field $v_0$ by requiring that
it shall equal $X_\epsilon$ in $M\setminus U_\epsilon(\cg)$,
and that for $x$ in $U_\epsilon(\cg)$,
it shall be given by
$$v_0(x)\equiv (\PV\grad H)(x)+h(x)(\bar{Q}\grad H)(x) ,$$
where $h:U_\epsilon(\cg)\rightarrow[0,1]$ is a smooth function obeying
$h|_{\cg}=1$ and $h|_{\partial U_\epsilon(\cg)}=0$.
In particular, $h$ shall
be strictly monotonic along all non-constant trajectories of the flow
generated by $\PV\grad H$, and the one form $dh$ shall vanish on $\cg$.

It can now be easily verified that
$v_0$ possesses all of the desired properties.
That it generates a gradient-like flow can be seen from
the fact that outside of $U_\epsilon(\cg)$,
$g(\grad H,v_0)=g(\grad H,X_\epsilon)$ is strictly positive,
as has been shown in the proof of proposition
{~\ref{CZineqprp1nongeneric}}.
Inside of $U_\epsilon(\cg)$, one finds
$$g(\grad H,v_0)=\|\PV\grad H\|_g^2 +
      h\|\bar{Q}\grad H\|_g^2 ,$$
due to the $g$-orthogonality both of $\PV$ and $\bar{Q}$.
The first term on the right hand side vanishes everywhere on
$\cg$, but nowhere else in $U_\epsilon(\cg)$.
The second term reduces to $\|\bar{Q}\grad H\|_g^2$ on $\cg$.
Since evidently, $\bar{Q}\grad H|_{\cg}$ is the gradient field of the
Morse function
$H|_{\cg}:\cg\rightarrow\R$ relative to the Riemannian metric on $T\cg$
that is induced  by
$g$, its zeros are precisely the critical points of $H$ on
$\cg$, and there are no other zeros apart from those.
This shows that the right
hand side of the above expression is strictly positive except at the critical
points of $H$, where everything vanishes.
Because the scalar product of $v_0$ with  $\grad H$ is strictly
positive except at the critical points of $H$, it is clear that $-v_0$
generates a gradient-like flow $\psi_{0,t}$,
so that $H$ is strictly decreasing along
all non-constant orbits.
Furthermore, it is also clear from the given construction that
$v_0$ is tangent to $\cg$.

Next, we address the proof that the
zeros of $-v_0$ are hyperbolic,
and that the number of negative eigenvalues of the Jacobian matrix
at any zero equals the corresponding Morse index of $H$.

To this end, we pick a local chart at a critical point $a$ of $H$.
The Jacobian matrix of $v_0$ at $a$ in this chart is given by
\eqn{\rm Jac}_a(v_0)&=& {\rm Jac}_a(\PV\grad H)+{\rm Jac}_a(\bar{Q}\grad H)
      \nonumber\\
      &=& \IPa P_a (D^2_a H)^\sharp +\IbQa
     \bar{Q}_a (D^2_a H)^\sharp  \nonumber\\
\label{armageddon}
      &=& (D^2_a H)^\sharp +(\IPa P_a- \IQa Q_a)(D^2_a H)^\sharp  .\eeqn
To explain this result, let us first
of all notice that there is no dependency on the
function $h$ because $dh|_{\cg}$ is zero.
Furthermore, $(D^2_a H)^\sharp$ is defined
as the matrix $[g^{ij}H_{,jk}|_a]$ in the given chart,
and $P_a$ denotes the matrix of $\PV(a)$.
Linearizing the vector fields $\PV\grad H$ and $\bar{Q}\grad H$
at a critical point of $H$,
all terms involving first derivatives of
$H$ are zero. This explains the second line.
The third line simply follows
from $\bar{Q}_a={\bf 1}_{2n}-Q_a$.

Normal hyperbolicity follows from the invertibility of ${\rm Jac}_a(v_0)$,
which is a consequence of lemma {~\ref{SpecQaminPalm}} below.
\\

\noindent{\bf Definition of the homotopy of vector fields.}
Let
$$v_\sig \equiv \sig \grad H + (1-\sig )v_0  $$
with $\sig \in[0,1]$.
We claim that for arbitrary $\sig $, $-v_\sig $ generates a
gradient-like flow $\psi_{\sig ,t}$,
in a manner that $H$ is strictly decreasing
along all non-constant orbits.
Furthermore, we claim that the zeros of
$v_\sig $ are hyperbolic fixed points of $\psi_{\sig ,t}$
that do not depend on $\sig $.
It then follows that the dimensions of the unstable manifolds are, for all
$\sig $, given by the Morse indices of the critical points of $H$ from which
they emanate.

To prove these claims, we consider the scalar product
$$g(\grad H,v_\sig )=\sig \|\grad H\|_g^2 +(1-\sig )g(\grad H,v_0) .$$
The first term on the right hand side is obviously everywhere positive except
at the critical points of $H$, and the same fact has been proved above for the
second term.
Thus, $H$ is strictly decreasing along all non-constant orbits of
$\psi_{\sig ,t}$, which proves that it is gradient-like.

The Jacobian of $v_\sig $ at a critical point of $H$ is given by
\eqnn{\rm Jac}_a(v_\sig )&=&\sig  (D^2_a H)^\sharp  +
      (1-\sig ){\rm Jac}_a(v_0)
    \\ &=&(D^2_a H)^\sharp +(1-\sig )(\IPa P_a - \IQa Q_a)(D^2_a H)^\sharp \\
      &=&({\bf 1}_{2n}+(1-\sig )(\IPa P_a - \IQa Q_a))(D^2_a H)^\sharp . \eeqnn
If we can prove that ${\rm Jac}_a(v_\sig )$ is invertible for all
$\sig \in [0,1]$, it follows
that the number of negative eigenvalues is independent of $\sig $.
To prove that this is indeed the case, we observe that because
$(D^2_a H)^\sharp$ is invertible,
one merely has to show that $({\bf 1}_{2n}+(1-\sig )(\IPa P_a - \IQa Q_a))$
is invertible.
This in turn is satisfied if $(\IPa P_a - \IQa Q_a)$ has no eigenvalues in
$[1,\infty)\subset\R$.

\begin{lm}\label{SpecQaminPalm}
The spectrum of
$\IPa P_a - \IQa Q_a$ has empty intersection with $[1,\infty)$.
\end{lm}

\prf
By arguing by contradiction.
Since $\IPa P_a - \IQa Q_a$ is selfadjoint with respect to $g_a$, it
is diagonalizable, and has a real spectrum.
Let us assume that $\kappa\in[1,\infty)$
is an eigenvalue
associated to the eigenvector $w\in T_a M$, so that
\eqn\label{manna}(\IPa P_a - \IQa Q_a)w=\kappa w .\eeqn
We multiply both sides of the equality sign with $P_a \IQa Q_a$ from the left,
and get the equation
$P_a \IQa Q_a \IPa P_a w = (1-\kappa) P_a \IQa Q_a w $.
On the other hand, multiplication from the left with $P_a$ gives
$P_a\IQa Q_a w = (1+\kappa) P_a w $.
Combining these two results, one obtains
$P_a \IQa Q_a\IPa P_a w = (1-\kappa^2) P_a w $,
which shows that $P_a w$ is an eigenvector of
$P_a\IQa Q_a\IPa P_a :V_a\rightarrow V_a$
that belongs to the eigenvalue $(1-\kappa^2)$.
We have proved earlier that normal hyperbolicity of $\cg$ implies that
$P_a\IQa Q_a\IPa P_a$ is invertible on $V_a$.
In addition, it is evident that $(1-\kappa^2)\leq 0$.

Let us assume that $P_a w\neq 0$. Then,
$$g_a(w,P_a\IQa Q_a\IPa P_a w) = g_a(Q_a\IPa P_a w,Q_a\IPa P_a w)>0$$
follows from the $g$-orthogonality of the projectors.
On the other hand, we also have
$$g_a(w,P_a\IQa Q_a\IPa P_a w) = (1-\kappa^2)g_a(w,P_a w) =
       (1-\kappa^2)g_a(P_a w,P_a w)\leq 0 ,$$
which is a contradiction. Hence, $P_a w =0$. In this case,
(~\ref{manna}) reduces to $\IQa Q_a w = \kappa w$, and multiplication with
$Q_a \IPa P_a$ from the left gives
$Q_a\IPa P_a \IQa Q_a w = \kappa Q_a\IPa P_a w = 0 $.
Because $Q_a\IPa P_a\IQa Q_a$ is invertible on $N_a\cg$, this
implies that $Q_a w$ is zero.
Therefore, $w$ is not contained in the
intersection of the images of $\IPa P_a$ and $\IQa Q_a$.
However, being an eigenvector that solves (~\ref{manna}),
it must be contained in this space, which is a contradiction.
\qed

The conclusion is that for all $\sig \in[0,1]$,
the zeros of $v_\sig $ are
hyperbolic fixed points of $\psi_{\sig ,t}$, and that the dimensions of the
corresponding unstable manifolds are given by the respective Morse indices of
$H$.
\\

Since
$\partial_t\psi_{\sig ,t}=-v_\sig $ depends smoothly on $\sig $,
hence $\psi_{\sig ,t}$ is $C^\infty$ in $\sig $.
Thus, $\sig$ smoothly
parametrizes a homotopy of stable and unstable manifolds emanating from the
critical points of
$H$, which belong to the gradient-like flow $\psi_{\sig ,t}$.

Because the fixed points of $\psi_{\sig ,t}$ are independent of $\sig $,
and
since the corresponding dimensions of the unstable manifolds coincide with
the
Morse indices of the critical points of $H$,
we again consider the free $\Z$-module
$$\C=\oplus_p \C^p$$
that is generated by the critical points of $H$, and graded by their Morse
indices.

For every fixed $\sig$, we define a coboundary operator on $\C$, using the
flow
$\psi_{\sig ,t}$. In fact,
picking a pair of critical points of
$H$ with a relative Morse index $1$, consider the unstable manifold
$W^-_{\sig ,a}$ of $a$, and the stable manifold
$W^+_{\sig ,a'}$ of
$a'$ associated to $\psi_{\sig ,t}$,
which are both smoothly parametrized by $\sig $.
Since $\sig $ parametrizes a homotopy of such manifolds,
they naturally
inherit an orientation from the one picked for
$\sig =1$ in the definition of the Morse-Witten complex for $(M,H)$.

Let
$\Sigma_E$ denote regular energy surface
for an arbitrary energy value $E$ between $H(a)$ and $H(a')$.
The intersection of $W_{\sig ,a}^\pm$ with any regular energy level surfaces
$\Sigma_E$ of $H$ is transverse,
because $H$ strictly decreases along all
non-constant orbits generated by $-v_\sig $.

$W^-_{\sig ,a}\cap\Sigma_E$ and
$W^+_{\sig ,a'}\cap\Sigma_E$ are oriented submanifolds of $\Sigma_E$,
and smoothly parametrized by $\sig $.
Hence, they define two homotopies of manifolds in $\Sigma_E$.
Their intersection number, being a homotopy invariant, is independent of
$\sig $, hence it equals the value obtained in case of $\sig =1$.
This implies that the coboundary operators obtained for arbitrary
$\sig $ are identical to the $\delta$-operator of the Morse-Witten complex given
for $\sig =1$, since their respective matrix elements are equal.
\\

To finally prove (~\ref{wonder}), we use the fact that
all stable and unstable manifolds of $\psi_{0,t}$ are, by construction,
either confined to some $\crit_i$, or otherwise, that they connect critical
points lying on different $\crit_i$'s.
This is because the stable and unstable
manifolds of the flow generated by $\PV\grad H$ only connect different
connectivity components of $\cg$.

Let us next consider pairs of critical points of $H$ with a relative Morse
index $1$ that lie on the same component
$\crit_i\in\cg$, and the corresponding stable and unstable manifolds of
$\psi_{0,t}$ which are contained in $\crit_i$.
Since
$v_0|_{\crit_i}$ simply is the projection of $\grad H|_{\crit_i}$ to
$T\crit_i$, these stable and unstable manifolds are precisely those which
were used to define the Morse-Witten complex on
$(\crit_i,H_i)$.

Picking only the stable and unstable manifolds of $\psi_{0,t}$
contained in $\cg$, we construct an operator $\tilde{\delta}$
acting on $\C$ in the
same manner in which the coboundary operator was defined. It is again a
coboundary operator, but now it is given by
$$\tilde{\delta}\equiv \oplus_i \delta_i .$$
$\delta_i$ denotes the coboundary operator of the Morse-Witten complex
associated to the pair $(\crit_i,H_i)$.

Finally, we denote by $P_i:\C\rightarrow\C_i$
the projection of the free $\Z$-module $\C$
generated by all critical points of $H$ to the one generated by
those critical points which lie in $\crit_i$.
The above construction makes it evident that
removing all integral lines of $-v_0$ that connect critical points on
different connectivity components of $\cg$,
one arrives at $\delta_i=P_i\delta P_i$, so that
$$\tilde{\delta}=\oplus_i P_i\delta P_i $$
(note that $\delta$ can be written as $\delta=\oplus_i\delta P_i$).
Inclusion of the missing integral lines 
would yield $\delta$, as our homotopy argument has proved.
This immediately makes the inequality
$${\rm dim} ({\rm im}\delta|_{\C^p})\geq{\rm dim}
     ({\rm im}\tilde{\delta}|_{\C^p})$$
clear.
We observe that this is precisely what is expressed in
the inequality (~\ref{wonder}). \qed

\pagebreak
\section{\nbf STABILITY CRITERIA FOR FIXED POINTS}

In this section, we will be interested in formulating
stability criteria for
equilibrium solutions of the constrained Hamiltonian system
$(M,\omega,H,V)$. In case of exponential (in)stability, the
discussion is elementary, and the results are standard.

However, in the marginally stable case, in which the linearized
dynamics is oscillatory in nature, it is much harder to arrive
at stability criteria if $V$ is not integrable (the integrable
case is again not interesting for us).

We will give a heuristic
line of arguments that supports a certain stability criterion
for this situation. It will be obtained from an elementary
application of averaging theory, and involves an incommensurability
condition imposed on the frequencies defined by the linearized
problem. Using a perturbation expansion that is adapted to the
flag of $V$, we will argue why this condition, which could merely be
an artefact of the averaging method, can presumably not be
dropped.

A rigorous proof of the conjectured
stability criterion is far beyond the scope of this text, and
is presumably at least as hard as proofs in KAM and
Nekhoroshev theory.
\\

\subsection{\nit Results of averaging}

For the discussion of stability, we again recall use
auxiliary K\"ahler metric $g$ on $M$, and
denote the induced Riemannian distance function by $dist_R$ (in contrast to
the Carnot-Caratheodory distance function $d_{C-C}$
induced by $g$, which will be considered
later).

\begin{dfi}\label{stabilitydfi}
A point $x_0\in\crit$ is stable if there
exists $\delta(\epsilon)>0$ for every $\epsilon>0$, so that for all $t$,
$dist_R(\tPhi_t(x),x_0)< \epsilon$
holds for all $x$ with $dist_R(x,x_0)<\delta(\epsilon)$.
\end{dfi}

Let $x_0\in\cg$,
and pick some small neighborhood
$U(a)\subset M$ together with an associated Darboux chart,
with its origin at $x_0$.
The equations of motion are given by
\eqn\label{Darbcheqsmoconstr}\partial_t x\;=\;P(x){\cal J} H_{,x}(x)
    \;=\; X_H^V(x),\eeqn
where $x=(x^1,\dots,x^n,x_{n+1},\dots,x_{2n})$,
and ${\cal J}$ is the symplectic standard matrix.
Furthermore, $H_{,x}$ abbreviates $\partial_{x}H$,
and $P$ is the matrix of the projector
$\PV$. $\omega$-orthogonality of $\PV$ translates into
$P(x){\cal J}X(x)={\cal J}P^\dagger(x)X(x)$ for all vector
fields $X$.

We will next bring the equations of motion in the vicinity of $0$,
corresponding to the point $x_0$, into
standard form. To this end, let
$$u:=\bar{P}_0 x\hspace{2cm}\tilde{y}:= P_0 x$$
denote new coordinates, picked to be
mutually orthogonal with respect to the K\"ahler
metric $g|_0$ in $\R^{2n}\cong T_0M$.

\begin{lm}\label{cggraphFlm}
Locally, $\cg$ is the graph of a $C^2$ function
$F:\bar{P}_0\R^{2k}\rightarrow\R^{2k}$, $u\mapsto F[u]$.
\end{lm}

\prf
The images of $P_0$ and $T_0\cg$ together
span $T_0M\cong\R^{2n}$.
To see this, let $\bar{Q}_0$ again denote the orthoprojector
$T_0 M\rightarrow T_0\cg$. The claim is then implied by the fact
that the matrix
$\IP P_0+\IbQ \bar{Q}_0=\1_{2n}+(\IP P_0-\IQ Q_0)$ is invertible.
The latter has already been proved above in the
discussion of the induced Morse-Witten complexes on $\crit_i$.
Thus, the rank of
$\bar{P}_0\IbQ\bar{Q}_0=\bar{P}_0(\IP P_0+\IbQ\bar{Q}_0)$ equals
the rank of $\bar{P}_0$, namely $2(n-k)$, hence
$\bar{P}_0$ projects $T_0M$ surjectively onto $T_0\cg$.
Consequently,
there is  a linear map
$G:\bar{P}_0\R^{2k}\rightarrow\R^{2k}$ for which
$T_0M$ is the graph $(u,G(u))$. $\cg$ thus admits a local
parametrization $(u,F[u])$ with $F(u)=G(u)
+O(\|u\|^2)$, for sufficiently small $\|u\|$, which is
$C^2$ since $\cg$ is assumed to be smooth.
\qed

\begin{prp}
The equations of motion are, in the coordinates $(y,z)$, given by
\eqn\partial_t y(t)&=&\Al
    \;y(t) \;+\;Y[z(t),y(t)]\\
    \partial_t z(t)&=&Z[z(t),y(t)],\label{parttzZeq}\eeqn
where $|Y[z,y]|,|Z[z,y]|=O(\|y\|\;\|z\|)+O(\|y\|^2)$,
and $\Al:=P_0 DX_H^V[0,0] \IP$.
\end{prp}

\prf
We introduce the function
$y[\tilde{y},u]:=\tilde{y}-F[u]$, which defines the coordinate transformation
$\Phi:(u,\tilde{y})\mapsto (z,y)$.
The Jacobi matrix of its inverse
$\Phi^{-1}$ at $(z,y)$ is
\eqnn D\Phi^{-1}[z,y]\;=\;\left[\begin{array}{cc}
      \bar{P}_0+DF[u(z,y)]\bar{P}_0&0\\ -DF[u(z,y)]\bar{P}_0&P_0
    \end{array}\right]\; .\eeqnn
From the definition of $F$ follows that $P_0 DF[u]=DF[u]$.
Thus, the equations of motion are now represented by
\eqn\partial_t z&=&\left(\bar{P}_0\;+\;DF[u(z,y)]
     \bar{P}_0\right) \;X[z,y]\eeqn
and
\eqn\partial_t y&=&\left(P_0\;-\; DF[u(z,y)]
      \bar{P}_0\right)\;X[z,y]\; ,\eeqn
where $X[z,y]\equiv X_H^V(\Phi^{-1}(z,y))$.
In this chart, $(z,0)$ parametrizes $\cg$, thus,
by definition of $\cg$, $X[z,0]=0$ for all $z$.
Taylor expansion of $\Phi^{-1}$ relative to $(0,0)$ gives
$$\Phi^{-1}[z,y]\;=\;\Phi^{-1}[0,0]\;+\;D\Phi^{-1}[0,0]\;
    \left(\begin{array}{c}z\\y\end{array}\right)+O(\|x\|^2).$$
Because of $F(0)=0$, $\Phi^{-1}(0,0)=(0,0)$.
Taylor expansion of $X_H^V(u,\tilde{y})$ relative to $(0,0)$ yields
$$X_H^V[y,z]\;=\;DX_H^V[0,0]\;\left(\begin{array}{c}u\\ \tilde{y}
    \end{array}\right)
    +\tilde{R}[u,\tilde{y}]\; ,$$
where $\tilde{R}[u,\tilde{y}]$ is a quadratic remainder term.
Consequently,
$$X[z,y]\;=\;DX_H^V[0,0]\;D\Phi^{-1}[0,0]
    \left(\begin{array}{c}z\\y\end{array}\right)+R[y,z]$$
with a quadratic remainder term $R[y,z]$.
It follows from $\bar{P}_0 DX_H^V(0,0)=0$ that (~\ref{parttzZeq})
holds for 
\eqn\label{wings}Z[z,y]&=&\left(\bar{P}_0\;+\;
     DF[u(z,y)]\bar{P}_0\right)
     R[z,y]\;.\eeqn
Furthermore,
\eqn\partial_t y&=&P_0\; DX_H^V[0,0]\;\IP\; y\nonumber\\
    &&-\;P_0\;DX_H^V[0,0]\;
    DF[u(z,y)]\;\left(\begin{array}{c}z\\y\end{array}\right)
    \;+\;\tilde{Y}[z,y] ,\eeqn
where $\tilde{Y}[z,y]$ is a quadratic remainder terms.
Finally, the kernel of $DX_H^V[0,0]$ is the tangent space
$T_0\cg$, which is also the image of $DF[0]$, thus
$DX_H^V[0,0]DF[0]=0$.
Thus,
\eqn\label{Darbcheqsmoconstrola}\partial_t y\;=\;P_0\; DX_H^V[0,0]\;
    \IP\; y
    \;+\;Y[z,y] ,\eeqn
with a quadratic Taylor remainder term $Y[z,y]$.
Clearly, $DX_H^V[0,0]={\cal J}\IP A_0$.

$\cg\cap U(x_0)$ is a center
manifold for the local system of ordinary differential equations
(~\ref{Darbcheqsmoconstrola}) and (~\ref{wings}),
which is parametrized by $(z,y=0)$. \qed

\noindent{\bf Asymptotic (in)stability.}
A standard application of the center manifold theorem shows that
if the spectrum of
$\Al$ does not intersect $i\R$,
there is a coordinate transformation
$(y,z)\rightarrow(\bar{y},\bar{z})$, so that  (~\ref{Darbcheqsmoconstrola}) and
(~\ref{wings}) can be written as
\eqn\partial_t\bar{y}(t)&=&\Al\; \bar{y}(t)
       \;+\;\bar{Y}[\bar{y}(t),\bar{z}(t)]\nonumber\\
         \partial_t\bar{z}(t)&=&0  \eeqn
\cite{ZeBlMa}, where $\bar{Y}(0,\bar{z})=0$ for all $\bar{z}$.
Thus, $x_0\in\cg$ is asymptotically unstable if
there are eigenvalues with a positive real part, and asymptotically stable if
all eigenvalues have a negative real part.

In case of integrable $V$, asymptotic stability is impossible,
because the eigenvalues always
come in pairs or quadruples with both positive and negative real parts.
However, if $V$ is nonintegrable, there is to the author's knowledge no
obstruction to the existence of asymptotically stable
equilibria, as the flow map is not symplectic.
\\

\noindent{\bf Marginal stability.}
The case of marginal stability is given
when spec$(\Al)\subset i\R\setminus\lbrace0\rbrace$.

We will in this case conjecture a stability criterion, which is motivated
by heuristic results obtained from
averaging theory on the one hand, and from a perturbation expansion that is adapted
to the flag of $V$, on the other hand.
A rigorous stability analysis is beyond the scope of this text.

To this end, let us assume that
$\Al$ is diagonalizable over
${\bf C}$.
To diagonalize it, we assume
that the  vector fields $Y[y,z]$ and $Z[y,z]$ in  (~\ref{Darbcheqsmoconstrola}) and
(~\ref{wings}) are analytic in $(y,z)$,
so that they possess a unique analytical continuation
into a complex vicinity of $x_0\in\cg$.
The system of ordinary differential equations
(~\ref{Darbcheqsmoconstrola}) and
(~\ref{wings}) is then interpreted in the sense that
$(y,z)$ is a vector in
$\C^{2k}\times\C^{2n-2k}$ (of small norm).
The continuation of $\cg$ into $\C^{2n}$ is defined by the common
zeros of $Y[0,z]$ and $Z[0,z]$ for $z\in\C^{2(n-k)}$.

Let spec$(\Al)=\lbrace i\omega_1,\dots,i\omega_{2k}
\rbrace$, with $\omega_i\in \R$.
There exists a linear transformation
$$\Psi:\C^{2n}\rightarrow\C^{2n}\;,$$
such that $\Al$ is diagonal in the new coordinates,
which, by abuse of notation, we denote again by $(y,z)$.
The equations of motion then reduce to
\eqn\label{beginning}\partial_t y(t)&=& {\rm diag}(i\omega) y(t)\;
         +\;Y[y(t),z(t)]\nonumber\\
         \partial_t z(t)&=&Z[y(t),z(t)]\;, \eeqn
where $\omega$ denotes the vector $(\omega_1,\dots,\omega_{2k})$.

Let us next introduce polar coordinates $(I,\phi)$ and $(J,\theta)$ in terms of
$$y^r =: e^{i\phi_r} I^r\hspace{2cm}z^s=: e^{i\theta_s} J^s$$
with $r=1,\dots,2k$ and $s=1,\dots,2n-2k$. In particular,
$I\in\R^{2k}$, $J\in\R^{2n-2k}$,
$\phi\in[0,2\pi]^{2k}=\TT^{2k}$ (the $2k$-dimensional torus), and
$\theta\in[0,2\pi]^{2n-2k}=\TT^{2n-2k}$. For brevity, vectors $(e^{i\phi_r}
v^r)$ and
$(e^{i\theta_s} w^s)$ will be denoted by $e^{i\phi} v$ and
$e^{i\theta} w$, respectively, where
$v\in\R^{2k}$ and
$w\in\R^{2n-2k}$.

In polar coordinates, the complexified equations of motion
for $\dot{y}$
(the dot abbreviates $\partial_t$) are given by
\eqn e^{i\phi}\dot{I}+i\dot{\phi}e^{i\phi}I = {\rm diag}(i\omega)
     e^{i\phi}I + Y[e^{i\phi}I,e^{i\theta}J],\eeqn
so that
\eqn\dot{I} &=& {\rm Re}\lbrace e^{-i\phi}
           Y[e^{i\phi}I,e^{i\theta}J]\rbrace\nonumber \\ \label{castor}
       \dot{\phi}&=&\omega + {\rm Im}\lbrace e^{-i\phi}
       {\rm diag}(\partial_I) Y[e^{i\phi}I,e^{i\theta}J]\rbrace . \eeqn
In the same manner,
\eqn \dot{J} &=& {\rm Re}\lbrace e^{-i\theta}
           Z[e^{i\phi}I,e^{i\theta}J]\rbrace \nonumber\\ \label{pollux}
       \dot{\theta}&=&{\rm Im}\lbrace e^{-i\theta}
       {\rm diag}(\partial_J) Z[e^{i\phi}I,e^{i\theta}J]\rbrace . \eeqn
$(I,J)\in\R^{2n}$ lies in a small vicinity of the origin.

Next, we fix a small parameter $\epsilon:=\|I(0)\|$,
and require that $\|J(0)\|\leq
O(\epsilon^2)$. Then, we redefine the variables
$I\rightarrow\epsilon I$ and $J\rightarrow \epsilon^2 J$ by rescaling. The new
coordinates $(I,J)$ have a norm of the order $O(1)$.

Analyticity of $Y[y,z]$ and $Z[y,z]$ in $(y,z)$ implies that the
power series expansion with respect to
$e^{i\phi}I$ and $e^{i\theta}J$ converges
for all $(I,J)\in\R^{2n}$ sufficiently close
to the origin.
In this manner,
(~\ref{castor}) and (~\ref{pollux}) yield
\eqn \label{bog1}
     \dot{I}^r&=&\sum_{|m|+|p|\geq 2}\epsilon^{|m|+2|p|-1}F^r_{mp}(I,J)
       e^{i(\langle m,\phi\rangle-\phi_r)}
      e^{i\langle p,\theta\rangle}\\ \label{bog2}
     \dot{J}^s&=&\sum_{|m|+|p|\geq 2}\epsilon^{|m|+2|p|-2}G^s_{mp}(I,J)
       e^{i\langle m,\phi\rangle}e^{i\langle p,\theta\rangle} \\ \label{bog3}
     \dot{\phi}_r&=&\omega_r +
       \sum_{|m|+|p|\geq 2}\epsilon^{|m|+2|p|-1}\Phi_{r;mp}(I,J)
       e^{i(\langle m,\phi\rangle-\phi_r)}
       e^{i\langle p,\theta\rangle}\\ \label{bog4}
     \dot{\theta}_s&=&
       \sum_{|m|+|p|\geq 2}\epsilon^{|m|+2|p|-2}\Theta_{s;mp}(I,J)
       e^{i\langle m,\phi\rangle}e^{i\langle p,\theta\rangle} ,\eeqn
introducing the multiindices $m\in\Z^{2k}$ and $p\in\Z^{2n-2k}$, with
$|m|:=\sum|m_r|$ and $|p|:=\sum|p_s|$. Of course, the right hand sides here are simply
the Fourier expansions with respect to
the $2\pi$-periodic angular variables
$\phi$ and $\theta$. Every Fourier coefficient, labelled by a pair of
indices $(m,p)$, is a homogenous polynomial of degree $|m|$ in  $I$,
and of degree $|p|$ in $J$.

In the limit $\epsilon\rightarrow 0$,
\eqn
\dot{I}=0\hspace{1.5cm}\dot{J}=0\hspace{1.5cm}\dot{\phi}=\omega\hspace{1.5cm}
     \dot{\theta}=0 \;.\eeqn
Under the assumption that all components of $\omega$ are rationally independent,
averaging can be applied with respect to the variable $\phi$ in order to
obtain an approximation to the long time behaviour of the
perturbed system.

The new, "averaged" variables are obtained via
$$f_t(\phi)\longrightarrow\bar{f}_t\;:=\;
       \frac{1}{(2\pi)^n}\int_{\TT^{n}}d^n\phi\;
       f_t(\phi) .$$
If the frequencies $\omega_r$ satisfy certain
incommensurability conditions, it can be proved that
the long time
behaviour of the system is very well approximated by the averaged solutions.

The only quantities in (~\ref{bog1}) $\sim$ (~\ref{bog4}) that do not vanish
under averaging
with respect to $\phi$ are the corresponding zero mode Fourier coefficients.

We recall that in the initial equations of
motion (~\ref{beginning}), the functions $Y[y,z]$ and $Z[y,z]$ are at least
$O(\|y\|)$. Thus, the leading terms of their power series in $(y,z)$ are at least
homogenous of degree $1$ in $y$, and thus involve terms
$e^{i\langle m,\phi\rangle}$ with $|m|\geq 1$, but no terms with $|m|=0$. A
quick inspection of the right hand sides of (~\ref{castor}) and
(~\ref{pollux}) assures us of the fact that (~\ref{bog1}) and (~\ref{bog3})
yield terms that survive the averaging process, but not (~\ref{bog2}) and
(~\ref{bog4}). Therefore, averaging the perturbed equations of motion
(~\ref{bog1}) $\sim$ (~\ref{bog4}) with respect to $\phi$ gives
\eqn \dot{\bar{I}}=\epsilon^2 \tilde{F}(\bar{I},\bar{J},\bar{\theta})
      \hspace{1.5cm}\dot{\bar{J}}=0\hspace{1.5cm} \dot{\bar{\theta}}=0 \eeqn
for some function $\tilde{F}$, where the bars account for averaged variables.

Returning to the initial coordinate chart for real $(y,z)$,
and the notation $x_0$ for the fixed point in discussion,
let us now assume that the quadratic form on $V_{x_0}$ defined by
$D_{x_0}^2 H|_{V_{x_0}}$ is positive
definite. We claim that the averaging result then suggests that $x_0$ is
stable. To this end, we notice that Taylor expansion of the
Hamiltonian (which is an integral of motion for the constrained system)
relative to $x_0$ gives
$$H(x)\;=\;H(x_0)\;+\;u^i\;\partial_{x^i} H(x_0)\; 
        \;+\;\frac{1}{2}\;y^i\;y^j\;\partial_{x^i}\;\partial_{x^j}
        H(x_0) \;+\; O(\epsilon^3),$$
and recall that $u=\bar{P}_{x_0} x$.
Because of lemma {~\ref{cggraphFlm}} the assumption
$\|z\|\leq O(\epsilon^2)$ implies that $\|u\|\leq O(\epsilon^2)$.
The quantity
$H(x)-H(x_0)=O(\epsilon^2)$ is an integral of motion; thus, if the
quadratic form
on $V_{x_0}$ defined by
$D_{x_0}^2 H$ is positive definite, $\|y\|$
has an order of magnitude $O(\epsilon)$ as long as $\|u\|=O(\epsilon^2)$
remains valid. The averaged equations of motion imply that
$\|\bar{z}\|=\|\bar{J}\|=O(\epsilon^2)$ is time-independent, so that due to
$\bar{u}=\bar{P}_0 \bar{z}$, the same applies to
$\|\bar{u}\|$.

In conclusion, we conjecture the following stability criterion.

\begin{conj}\label{stabcriteriaconj}
Let $x_0\in\cg$, {\rm spec}$(\Al)=\lbrace i\omega_1,\dots,
i\omega_{2k}\rbrace$, with $\omega_i\in\R\setminus\lbrace0\rbrace$. 
Assume that (1)
the frequencies $\omega_r$ are rationally independent, and (2) that
the quadratic form on $V_{x_0}$ defined by $D_{x_0}^2 H|_{V_{x_0}}$ is
    positive definite.
Then, $x_0$ is stable.
\end{conj}

\subsection{Dynamics along the flag of V}

Let us now approach the discussion of marginal stability from
a different angle. In the subsequent paragraphs, we
develop a geometrically invariant
description of the local dynamics in the
vicinity of $\crit$ which is adapted to the flag of $V$.
Our motivation is to comment on the origin of the
incommensurability condition imposed on the frequencies
involved in the above stability criteria, and to give arguments why
it can not be dropped.

Let us assume that $x$ lies in a small open
neighborhood $U$ of $x_0\in\crit$, and
that $\cg:=\crit\cap U$ satisfies the genericity
condition of theorem {~\ref{Sardthm}}.

\begin{prp}\label{tubneigh}
Let  $\cg=\crit\cap U$ have the genericity property
formulated in theorem {~\ref{Sardthm}}.
Then, there exists $\epsilon>0$ such that
every point $x\in U$ with $d_R(x,\cg)<\epsilon$
is given by
$$x=\exp_s Y (x_0)\;\;\;\;,\;\;\;\;|s| < \epsilon$$
for some $Y\in\Gamma(V)$ with $\|Y\|_{g_M}\leq 1$,
$x_0\in\cg$
($\exp_s Y$ denotes the 1-parameter group of diffeomorphisms
generated by $Y$, with $\exp_0 Y=$id).
\end{prp}

\prf
We pick a spanning family $\lbrace Y_i\in\Gamma(V)\rbrace_1^{2k}$ of $V$,
with $\|Y_i\|_{g_M}= 1$.
If for all $x_0\in\cg$, $T_{x_0}\cg$ contains no subspace of $V_{x_0}$, then
$$\exp_1(t_1Y_1+\dots+t_{2k}Y_{2k})(\cg)\;\;\cap\;\;U$$
is an open tubular neighborhood of $\cg$ in $U$,
for $t_i\in (-\epsilon,\epsilon)$.
Because the normal space $N_{x_0}\cg$ is dual to the span of
the 1-forms $dF_i$ at $x_0$,
this condition is satisfied if and only if the matrix
$[dF_j(Y_i)]=[Y_i(Y_j(H))]$
is invertible everywhere on $\cg$.
According to proposition {~\ref{invert}}, this condition
is indeed fulfilled. \qed

Due to proposition {~\ref{tubneigh}}, there is an element
$Y\in\Gamma(V)$ with $\|Y\|_{g_M}\leq 1$,
so that
$$x\;=\;\Psi_\epsilon(x_0)\;$$
for some $0<\epsilon\ll 1$.
Since $x_0\in\cg$, it is clear that under the flow
generated by $X_H^V$, $\tPhi_{\pm t}(x_0)=x_0$, thus
the solution of (~\ref{eqsofmo}) belonging to the initial condition $x$
is given by
\eqnn \Psi_\epsilon^t(x_0)\;:=\;
      \tPhi_t\circ\Psi_\epsilon(x_0)\;=\;\left(\tPhi_t\circ
      \Psi_{\epsilon}
      \circ\tPhi_{-t}\right)(x_0)\;.\eeqnn
In particular, $\Psi_\epsilon^t$ is the 1-parameter group of diffeomorphisms
with respect to the variable $\epsilon$ that
is generated by the pushforward vector field
\eqn\label{pushf}Y_t(x)\;:=\;\tPhi_{t\;*}\;Y(x)\;=\;
      d\tPhi_{t}\circ Y(\tPhi_{-t}(x))\;,\eeqn
where $d\tPhi_t$
denotes the tangent map associated to $\tPhi_t$.
This is a standard fact of differential geometry, cf. for instance
\cite{Jo}.
From the group property $Y_{s+t}=\tPhi_{s\,*} Y_t$ follows that
\eqn\label{Ytder}\partial_t Y_t\;=\;\left.\partial_s\right|_{s=0}
      \tPhi_{s\,*} Y_t \;=\;[X_H^V,Y_t]\;\eeqn
holds everywhere in $U$.

We recall from the beginning of the previous section that there
exists a local spanning
family  $\left\lbrace Y_i\in\Gamma(V)\right\rbrace_{i=1}^{2k}$
for $V$ that  satisfies
$$\omega(Y_i,Y_j)\;=\;\tJ_{ij}\;,$$
with
$\tJ:=\left[\begin{array}{cc}0&\1_k\\-\1_k&0\end{array}\right]$.
Furthermore, defining $\theta_i(\cdot):=\omega(Y_i,\cdot)$,
$$\PV\;=\;\tJ^{ij}Y_i\otimes \theta_j\;,$$
where $\tJ^{ij}$ are the components of $\tJ^{-1}=-\tJ$. Finally,
\eqnn X_H^V\;=\;\PV(X_H^V)\;=\;-\;Y_i(H)\;\tJ^{ij}\; Y_j\;\eeqnn
in the basis $\left\lbrace Y_i\right\rbrace_{i=1}^{2k}$.

\begin{prp}
Let $f,F_i\in C^\infty(U)$, where $F_i := Y_i(H)$, $i=1,\dots,2k$,
and assume that $F_i(\Psi_\epsilon^t(x_0))$,
$f(\Psi_\epsilon^t(x_0))$ are real analytic in $\epsilon$.
For $X,Y\in \Gamma(TM)$,
let $$\cL_Y^r X \;=\;[Y,\dots,[Y,X]]$$ denote the $r$-fold iterated
Lie derivative. Then, for sufficiently small $\epsilon$,
\eqn\label{dertflag}\partial_t f(\Psi_{\epsilon}^t(x_0))\;=\;-\;
       F_i(\Psi_\epsilon^t(x_0))\;\;\tJ^{ik}\;\;\sum_{r\geq 0}\;\;
      \frac{\epsilon^r}{r!}\;
      (\cL_{Y_t}^{\;r} Y_k)\;(f\circ\Psi_\epsilon^t)(x_0)\;.\eeqn
\end{prp}

\prf
Clearly,
\eqn\partial_t f(\Psi_\epsilon^t(x_0))&=&X_H^V(f)
     (\Psi_\epsilon^t(x_0))\nonumber\\
     &=&-\;
     F_i(\Psi_\epsilon^t(x_0))\;\tJ^{ik}\;
     Y_k(f)(\Psi_\epsilon^t(x_0))\nonumber\\
     &=&-\;
     F_i(\Psi_\epsilon^t(x_0))\;\tJ^{ik}\;
     \left(\Psi_{\epsilon\;*}^t\; Y_k \right)
     (f\circ\Psi_\epsilon^t)(x_0)\;.\eeqn
Using the Lie series
\eqn\Psi_{\epsilon\;*}^t\; Y_k\;=\;\sum_r\;\frac{\epsilon^r}{r!}\;
    \cL_{Y_t}^{\;r} Y_k\;,\eeqn
we arrive at the assertion. \qed

\begin{prp}\label{commflag}
Assume that $Y_{t=0}\in\Gamma(V)$, and let $\lbrace Y_j\rbrace_1^{2k}$
be the given local spanning family of $V$.
Then, $\cL_{Y_t}^{\;i}Y_j\in\Gamma(V_i)$, where $V_i$ is the
$i$-th flag element of $V$.
\end{prp}

\prf
Because of $\tPhi_{t\;*}:\Gamma(V)\rightarrow\Gamma(V)$, $Y_t$
is a section of $V$ for all $t$ if it is for $t=0$.
The claim immediately
follows from the
definition of the flag of $V$.
\qed

Proposition {~\ref{commflag}} implies that there are functions
$a^i(t,\cdot)\in C^\infty(U)$, $i=1,\dots,2k$, so that
\eqn Y_t(x)\;=\;a^i(t,x)Y_i\;.\eeqn
Their time evolution is governed by the following proposition.

\begin{prp}
Let $Y_{t=0}=a_0^i Y_i$ define the initial condition, and introduce
the matrix
$$\Omega_x:=[Y_l(F_i)(x)\tJ^{ij}]\;\;.$$
Then, pointwise in $x$,
\eqn\label{abODE} a^m(t,x)&=&\left(\exp(-\;t\;\Omega_x)\right)^m_j\;a_0^j
      \;+\;F_j(x)\;R^{jm}_i(t,x)\;a_0^i \;,\eeqn
where
$$R^{jm}_i(t,x)\;:=\; \tJ^{jl}\;\tJ^{nk} \int_0^t \;ds\;
     (\exp(-(t-s)\;\Omega_x))^m_k
      \;\omega\left([Y_l,\tPhi_{s\;*}Y_i]\;,\; Y_n\right)\;.$$
\end{prp}

\prf
The initial condition at $t=0$ is given by
$Y_0=a^i_0 Y_i$, that is, by $a^i(0,x)=a^i_0$.
Thus, by the definition of $Y_t$ in  (~\ref{pushf}),
one has $Y_t=a^i_0\, \tPhi_{t\;*}Y_i$,
so that
$$a^i(t,x)\;Y_i\;=\;a^i_0\; \tPhi_{t\;*}Y_i\;.$$
From $\omega(Y_i,Y_j)=\tJ_{ij}$, $\tJ_{ik}=-\tJ_{ki}$ and
$\tJ_{im}\tJ^{ml}=-\delta^l_i$,
$$a^l(t,x)\;=\;-\;a^i_0\;\omega\left(\tPhi_{t\;*}Y_i\;,\;Y_{j}\right)
     \;\tJ^{jl}\;.$$
Now, taking the $t$-derivative on both sides of the equality sign,
one finds
\eqnn\partial_t a^m(t,x)&=&-\;a^i_0\;
     \omega\left([X_H^V,\tPhi_{t\;*}Y_i]\;,\;Y_{k}\right)\;\tJ^{km}\\
     &=&-\;a^i_0\;(\tPhi_{t\;*}Y_i)(F_j)(x)\;\tJ^{jl}\;
      \omega\left(Y_l\;,\;Y_{k}\right)\;\tJ^{km}\nonumber\\
     &&
     -\;a^i_0\;F_j(x)\;\tJ^{jl}\;\tJ^{km}\;
      \omega\left([Y_l,\tPhi_{t\;*}Y_i]\;,\;Y_{k}\right)\\
     &=&-\;a^i(t,x)\;Y_i(F_j)(x)\;\tJ^{jm}\;\nonumber\\
     &&
     -\;a^i_0\;F_j(x)\;\tJ^{jl}\;\tJ^{km}\;
      \omega\left([Y_l,\tPhi_{t\;*}Y_i]\;,\;Y_{k}\right)\;.\eeqnn
Using the variation of constants formula pointwise in $x$, one
arrives at the assertion.
\qed

\subsubsection{Leading order perturbation theory}\label{Loptsubsubsect}

Let us next use the small parameter $\epsilon$
for perturbation theory. We will only be interested in a heuristic
argument that demonstrates why the incommensurability condition
on the eigenfrequencies in conjecture {~\ref{stabcriteriaconj}}
is presumably necessary.

The simplified case that we will consider is defined by the following
assumptions:
\newcounter{nn0}
\begin{list}
  {(\arabic{nn0})}{\usecounter{nn0}\setlength{\rightmargin}{\leftmargin}}
\item $\Omega_x=\Omega$, constant for all $x$ in $U$.
\item spec($\Omega$)$=\;\lbrace i\omega_1,\dots,i\omega_{2k}\rbrace$, with
   $\omega_r\in \R$.
\item $\|\;\Omega\;\|\;:=\;\sup_r|\;\omega_r\;|\;\ll\;\frac{1}{\epsilon}$.
\end{list}

Let us briefly comment on the generic properties
of $\lbrace\omega_r\rbrace$.
Writing $\Omega=\tJ A$,
we decompose the matrix $A=[Y_i(Y_j(H))(x_0)]$
into its  symmetric and antisymmetric parts
$A_+$ and $A_-$, respectively.
$A_-=[[Y_j,Y_i](H)(x_0)]/2$ vanishes
if $V$ is integrable, as one deduces from the fact that $X_H|_{x_0}$ is
a vector in $V_{x_0}^\perp$ for all $x_0\in \cg$,
and from the Frobenius condition.
The linear system of ODE's $\underline{\dot{a}}=\tJ A_+ \underline{a}$
in the space of $a^i(t)$'s
is Hamiltonian, hence the spectrum of $\tJ A_+$, if it is purely
imaginary, consists of complex conjugate pairs of eigenvalues in $i\R$
(here, we have introduced the notation $\underline{a}:=(a^1,\dots,a^{2k})$).
If $\tJ A_-$ admits a small relative norm bound with respect to
$\tJ A_+$, pairs of complex conjugate eigenvalues $\pm i\omega_r$ will generically
be deformed in a manner that they lose the property of
being identical up to sign.
Thus, generically,
we may assume that all frequencies $\omega_r$ are distinct from each other,
and there are as many negative as positive ones.

From (~\ref{abODE}), one infers
$$Y_t\;=\;a^j_0\; (\exp(-t\Omega))^i_j\; Y_i\;+\;\sum\; O(|x|)\;Y_i\;$$
because $|F_j(x)|=O(|x|)=O(\epsilon)$, since $F_j(x_0)=0$.

Thus,
$$[Y_t,X]\;=\;a^j_0 \exp(-t\;\Omega)_j^i\;[Y_i,X]\;+\;
      \sum\; O(\epsilon)\;[Y_i,X]
     \;+\;\sum\; O(1)\;Y_i $$
for all $X\in\Gamma(TM)$, and $x\in U_\epsilon(x_0)$.
Assuming that everything is sufficiently smooth,
iterating the Lie bracket $\cL_{Y_t}$ $r$ times produces
$$\left(\prod_{m=1}^r
       a^{j_m}_0\;\left(\exp(-t\;\Omega)\right)^{i_m}_{j_m}\;+\;O(\epsilon)
       \right)
       [Y_{i_1},[Y_{i_2},\dots,[Y_{i_r},Y_l]\cdots]]\;,$$
plus a series of terms with less than  $r$ nested Lie commutators,
which contribute to higher order corrections
(that is, $O(\epsilon^{r+1})$)  of
terms indexed by $r'<r$ in (~\ref{Vrcontr}).

Let us, for a discussion of leading order perturbation theory along each
flag element of $V$, drop the terms of order $O(\epsilon)$, and assume
that everything is sufficiently smooth so that our considerations
hold if
$t\leq O(\epsilon^{-1})$.

For fixed $r$,  let us consider the term
\eqn\label{Vrcontr}\;F_i(\Psi_\epsilon^t(x_0))\;
      \tJ^{ik}\;\left(\cL^{\;r}_{Y_t}Y_k\right) (f\circ \Psi_\epsilon^t)(x_0)\;,
      \eeqn
which describes the contribution of (~\ref{dertflag}) along the $r$-th flag
component of $V$, at least for sufficiently small $t$,.

To this end, we have
\eqn\label{Fiser} F_i(\Psi_\epsilon^t(x_0))\;=\;Y_t(F_i)(x_0)\;+
      \;O(\epsilon^2)\;,\eeqn
due to $F_i(x_0)=0$.
Therefore,
\eqn F_i(\Psi_\epsilon^t(x_0))\;\tJ^{ik}\;=\;\epsilon\;
     \exp\left(-t\Omega\right)^m_j\;a_0^j\;\Omega^k_m\;+\;
     O(\epsilon^2)\;,\eeqn
as a straightforward calculation shows.

Collecting all results obtained so far, the
terms with $r$ nested commutators in (~\ref{Vrcontr}) are
\eqnn\frac{\epsilon^{r+1}}{r!}\; a^i_0\;
       \exp(-t\;\Omega)^j_i\;\Omega_j^l\;\;\left(\prod_{m=1}^r
       a^{j_m}_0\; \left(\exp(-t\;\Omega)\right)^{i_m}_{j_m}
       \right)
       [Y_{i_1},[Y_{i_2},\dots,[Y_{i_r},Y_l]\cdots]](f)(x_0)\\
       \;+\;\;\;O(\epsilon^{r+2})\;,\eeqnn
as long as $dist_R(\Psi_\epsilon^t(x_0),x_0)\leq O(\epsilon)$.
This implies that for $f\in C^\infty(U)$,
\eqn f(\Psi_\epsilon^t(x_0))\;\approx\;f(x_0)\;+\;\sum_{r\geq 0}\;
       \frac{\epsilon^{r+1}}{r!}\;\int_0^t\;ds\; a^i_0\;
       \exp(-s\;\Omega)^j_i\;\Omega_j^l\;\times
       \hspace{2cm}\nonumber\\
       \times\;
       \left(\prod_{m=1}^r
       a^{j_m}_0\; \left(\exp(-s\;\Omega)\right)^{i_m}_{j_m}
       \right)
       [Y_{i_1},\dots,[Y_{i_r},Y_l]\cdots](f)(x_0)\;,
       \label{fflagexp}\eeqn
up to errors of higher order in $\epsilon$ for every fixed $r$,
as long as $dist_R(\Psi_\epsilon^t(x_0),x_0)\leq O(\epsilon)$.

If $f$ is picked as the $i$-th coordinate function $x^i$, so that
$f(\Psi_\epsilon^t(x_0))=x^i(t)$,
the quantity $[Y_{i_1},\dots,[Y_{i_r},Y_l]\cdots](f)(x_0)$ is the
$i$-th coordinate of the vector field defined by the brackets at $x_0$.
Consequently, (~\ref{fflagexp}) is the component decomposition of
$x^i(t)$ relative to the flag of $V$ at $x_0$, to leading order in
$\epsilon$.

By assumption for the simplified model,
$\Omega$ has a purely imaginary spectrum. In this case, the operator
$\exp(-s\Omega)$ has a norm 1 for all $s$. Consequently, the integrand
of (~\ref{fflagexp}) is bounded for all $s$. It follows that
if the integral should diverge, and become bigger than $ O(\epsilon)$,
it will take a time
$$t\geq O\left(\frac{1}{\epsilon^{\;r}}\right)$$
to do so along the flag element $V_r$. The leading term of order
$O(\epsilon)$, corresponding to $r=0$,
remains bounded for all $t$,
but higher order corrections to it might diverge.

Let us next write
\eqn\underline{a}(s)&=&\exp(-s\;\Omega)\;\underline{a}_0\nonumber\\
     &=&
     \sum_{\alpha=1}^{2k}\;
     A_\alpha\;\underline{e}_\alpha\;\exp(-i\omega_\alpha s)
     \;,\eeqn
where $\lbrace \underline{e}_\alpha\rbrace$ is a orthonormal
eigenbasis
of $\omega$ with respect to the standard
scalar product in $\C^{2k}$, and spec$(\Omega)=\lbrace i\omega_\alpha\rbrace$.
The amplitudes $A_\alpha\in \C$ are determined by the initial condition
$a^i(t=0)=a_0^i$, and will be assumed to be nonzero.
By linear recombination of the vector fields $Y_i$, one can set
$e^i_\alpha=\delta_{i,\alpha}$. Then,
(~\ref{fflagexp}) can be written as
\eqn\label{intnestcomm}\sum_{r\geq 0}\;\frac{\epsilon^{r+1}}{r!}\;
       \sum_{l;i_1,\dots,i_r}\; I_{l;i_1,\dots,i_r}(t)\;
        [Y_{i_1},\dots,[Y_{i_r},Y_l]\cdots](f)(x_0)\;,\eeqn
where
\eqn\label{Iindt}  I_{l;i_1,\dots,i_r}(t)&:=&
       \int_0^t\;ds\;
       \;\omega_l\; A_l\;
       \left(\prod_{m=1}^r
       A_{j_m}\;\right)\;\exp \left(-is\left(\omega_l\;+\;\sum_{m=1}^r
       \omega_{j_m}\right)\right)\nonumber\\
       &=&i\;\left(\omega_l\;+\;\sum_{m=1}^r
       \omega_{j_m}\right)^{-1}\;
       \;\omega_l\; A_l\;
       \left(\prod_{m=1}^r
       A_{j_m}\;\right)\;\times\nonumber\\
       &&\hspace{2cm}\times\;\left\lbrace\;
       \exp\left( -it\left(\omega_l\;+\;\sum_{m=1}^r
       \omega_{j_m}\right)\right)\;-\;1\;\right\rbrace\;\;.
       \label{Ili1dotsirdef}\eeqn
Evidently, the nested commutators vanish if all indices
$i_1,\dots,i_r,l$ have equal values.
\\

\subsubsection{Conditions for instability}

Let us introduce the set
\eqn\I^{(r)}(t)\;:=\;\left\lbrace I_{l;i_1,\dots,i_r}(t)
      \right\rbrace_{l,i_j=1}^{2k}\;
      \setminus\; \left\lbrace I_{l;l,\dots,l}(t)
      \right\rbrace_{l=1}^{2k}\;,\eeqn
which we endow with the norm
$$\|\I^{(r)}(t)\|\;:=\;\sup_{I(t)\in\I^{(r)}(t)}\; |I(t)|\;.$$
Furthermore, let
\eqn\|A\|\;:=\;\sup_{i=1,\dots,2k}\lbrace|A_i|\rbrace\;,\eeqn
where $A_i$ are $\C$-valued amplitudes.

We will now determine in which situations $\|\I^{(r)}(t)\|$ diverges
in the limit $t\rightarrow\infty$.

To this end, let
\eqn\Aa\;:=\;\lbrace\omega_1,\dots,\omega_{2k}\rbrace\eeqn
and denote the $r$-fold sumset by
\eqn\Aa_r\;:=\;\underbrace{\Aa\;+\;\cdots\;+\;
    \Aa}_{r\;{\rm times}}\;,\eeqn
defined by the set containing all sums of $r$ elements of $\Aa$.

For two sets of real numbers $\Aa$ and ${\mathfrak B}$, we define their
distance as
\eqn d(\Aa,{\mathfrak B})\;:=\;\inf_{i,j}\left\lbrace\;\left|a_i-b_j\right|\;|\;
     a_i\;\in\;\Aa\;,\;b_j\;\in\;{\mathfrak B}\;\right\rbrace\;.\eeqn
Then, it follows from (~\ref{Ili1dotsirdef}) that if $d(\Aa_r,-\Aa)>0$,
\eqn\|\I^{(r)}(t)\|\;\leq\;d(\Aa_r,-\Aa)^{-1}\;\|\Omega\|
     \;\|A\|^r\;\eeqn
(the sum over frequencies $\sum_{m=1}^r
\omega_{j_m}$ in (~\ref{Ili1dotsirdef}) is
an element of $\Aa_r$, and can only equal $-\omega_l$ if
$d(\Aa_r,-\Aa)=0$).
However, if $d(\Aa_r,-\Aa)=0$, there is a tuple of indices
$\lbrace l;i_1,\dots,i_r\rbrace$ such that
\eqn I_{l;i_1,\dots,i_r}(t)\;=\;-\;t\;\omega_l\;A_l\;\prod_{m=1}^r
       A_{j_m}\;.\eeqn
This is simply obtained by letting the sum of frequencies in
(~\ref{Ili1dotsirdef}) tend to zero. Thus, in this case,
\eqn\|\I^{(r)}(t)\|\;\sim\;t\;,\eeqn
that is, a divergence linear in $t$ as $t\rightarrow\infty$ (of
course, the validity of our leading order perturbation theory
breaks down as $t\rightarrow\frac{1}{\epsilon}$).
Only if there are simultaneously positive and negative frequencies,
$d(\Aa_r,-\Aa)=0$ is possible, but due to the remark
at the beginning of subsection {~\ref{Loptsubsubsect}},
this situation must generically
assumed to be given.

Let us consider some examples.
The fact that $\|\I^{(0)}(t)\|$ is bounded for all $t$ is trivial.
On the next level, $r=1$, we consider the component
in the direction of the first flag element $V_1=[V,V]$. The condition
for the emergence of a divergent solution is that $d(\Aa,-\Aa)=0$.
This is precisely given if there is a pair of frequencies $\pm\omega_i$
of equal modulus, but opposite sign. For $r=2$, assuming that
$d(\Aa,-\Aa)>0$, the condition $d(\Aa_2,-\Aa)=0$ implies that there
is a triple of frequencies such that $\omega_{i_1}+\omega_{i_2}=
-\omega_{i_3}$, $i_j\in\lbrace1,\dots,2k\rbrace$. If this occurs,
the solution will diverge in the direction of the second
flag element, $V_2=[V,[V,V]]$. The discussion for $r>2$ continues in
the same manner.

In conclusion, we have arrived at the following proposition.

\begin{prp}
If $d(\Aa_r,-\Aa)=0$ for some $r$, then $\|\I^{(r)}(t)\|= O( t)$ for
$t\rightarrow\infty$.
\end{prp}

The above results suggest that if the frequencies of the linearized
problem do not satisfy the incommensurability condition
"$d(\Aa_r,-\Aa)>0$ for all $r$", the equilibrium $x_0$ is unstable.
However, it has also become clear that the time required for an orbit to
leave a Riemannian
$\epsilon$-neighborhood $U_\epsilon(x_0)$ is extremely large.
Assume that $d(\Aa_r,-\Aa)=0$ for some $r\leq r(V)$
(the degree of non-holonomy of $V$). Then, it takes a
time on the order of $O(\frac{1}{\epsilon^{r}})$ to exit $U_\epsilon(x_0)$
along the flag element $V_r$,
according to our leading order
perturbative results (due to the factor $\frac{\epsilon^r}{r!}$ in
(~\ref{fflagexp})).

Although the perturbative results lose their validity
already after $t\leq O(\frac{1}{\epsilon})$,
it is always possible to continue solutions by picking new charts,
centered around new basis points after times
of order $O(\frac{1}{\sqrt{\epsilon}})$, say. There is no
doubt that since
$d(\Aa_r,-\Aa)=0$ holds for all basis points $x_0'\in U_\epsilon(x_0)$
(by assumption, $\Omega$ is constant in $U_\epsilon(x_0)$),
the same divergence will be observed in each chart, and the prediction that
$U_\epsilon(x_0)$ will indeed be exited after a
time of order $O(\frac{1}{\epsilon^{r}})$ is realistic. During this
time, the orbit will not drift away from $U_\epsilon(x_0)\cap\cg$
in the direction of $V_{x_0}$ if $D_{x_0}^2 H |_{V_{x_0}}$
is positive definite on $V_{x_0}$, as required in the conjectured
stability criterion. This suggests that the incommensurability
condition imposed on the frequencies of the linearized system
can indeed not be dropped.

We claim that although in the light of Riemannian geometry, divergences
of this type are completely unspectacular up to times of order
$\geq O(\frac{1}{\epsilon})$,
they are very severe in the light of the natural, 
internal geometry of the system,
which is not the Riemann, but the Carnot-Caratheodory structure 
induced by $g$.
\\

\subsubsection{Relation to sub-Riemannian geometry}

The structure exhibited in this discussion clearly
shows that the constrained Hamiltonian system
$(M,\omega,H,V)$ has many characteristics of systems usually encountered
in sub-Riemannian geometry \cite{BeRi,Ge,Gr2,Str}.

The natural metric structure that accounts for the particular structure of
the present system is obtained from
the Carnot-Caratheodory distance function
$dist_{C-C}$  induced by the Riemannian metric $g$.
It assigns to a pair of points
$x,y\in M$ the length of the shortest $V$-horizontal $g$-geodesic.

If $V$ satisfies the Chow condition, $dist_{C-C}(x,y)$ is
finite for all $x,y\in M$. This is the Rashevsky-Chow theorem \cite{BeRi,Gr2}.
In this case, the Carnot-Caratheodory $\epsilon$-ball
$$B^{C-C}_\epsilon(x_0)\;:=\;\left\lbrace\;x\in M\;\left|\;
       dist_{C-C}(x,x_0)\;<\;\epsilon\;\right\rbrace\right.$$
is open in $M$.

If $V$ does not satisfy the Chow condition, pairs of points
that cannot be joined by $V$-horizontal
$g_M$-geodesics are assigned a Carnot-Caratheodory distance $\infty$.
In this case, $M$ is locally foliated into
submanifolds $N_\lambda$
of dimension $(2n-{\rm rank}V_{r(V)})$,
with $\lambda$ in some index set,
which are integral manifolds of the
(necessarily integrable)
final element $V_{r(V)}$ of the flag of $V$
($r(V)$ denotes the degree of nonholonomy of $V$).
On every $N_\lambda$, the distribution
$V_\lambda:=j_\lambda^* V$ satisfies the Chow condition,
where $j_\lambda:N_\lambda\rightarrow M$ is the inclusion.
Therefore, all points $x,y\in N_\lambda$ have a finite
distance with respect to the Carnot-Caratheodory metric induced by the
Riemannian metric  $j_\lambda^*g_M$.
Every leaf $N_\lambda$
is an invariant manifold of the flow $\tPhi_t$.
\\

\noindent{\bf The ball-box theorem}.
Assume that the spanning family $\lbrace Y_{i_r}\rbrace_{r=1}^{r(V)}$
of $TM$ is suitably
picked so that $\lbrace Y_{i_r}\rbrace$ spans the flag element
$V_r$. Let the $g$-length of all $Y_{i_r}$'s be 1.
Then, we define the 'quenched' box
$${\rm Box}_\epsilon(x)\;:=\;\left.\left\lbrace
      \exp_1\left(\sum_{r=1}^{r(V)}
      \epsilon^r\sum_{i_r=1}^{{\rm dim}V_r}t_{i_r}\;Y_{i_r}\right)(x)\;
      \right|\;t_{i_r}\;\in\;(-1,1)\right\rbrace\;$$
in $N_\lambda$, where $\lambda$ is suitably picked so that $x\in N_\lambda$.
Evidently, if $V$ satisfies Chow's condition, $N_\lambda=M$.
According to the ball-box theorem \cite{BeRi,Gr2}, there are constants
$C>c>0$, such that
$${\rm Box}_{c\epsilon}(x)\;\subset\;B^{C-C}_\epsilon(x)\;
      \subset\;{\rm Box}_{C\epsilon}(x)\;.$$
Therefore, Carnot-Caratheodory $\epsilon$-balls can be approximated
by quenched boxes of Riemannian geometry.
\\

\noindent{\bf Instabilities in the light of C-C geometry.}
The above perturbative results imply that
if there is some $r<r(V)$, for which
$d(\Aa_r,-\Aa)=0$,
the flow $\tPhi_t$ blows up the quenched boxes,
and thus the Carnot-Caratheodory $\epsilon$-ball
around $x_0\in\cg$, linearly in $t$, and along the direction of $V_r$.
In fact,
$B^{C-C}_\epsilon(x_0)$ is widened
along $V_r$ at a rate linear in $t$. After a
time of the order $O(\frac{1}{\epsilon})$, the
Carnot-Caratheodory $\epsilon$-ball containing the initial condition
is increased to an extent that it can only be
embedded into a Carnot-Caratheodory
$O_\epsilon(1)$-ball.
In this light, the instabilities in discussion are very severe.

This type of instability has no counterpart
in systems with integrable constraints.

\pagebreak

\section{\nbf NONHOLONOMIC MECHANICS}

We will in this section focus on nonholonomic mechanical
systems, and their relationship to the constrained Hamiltonian
systems considered previously.
The discussion is restricted to Hamiltonian mechanical
systems that are subjected to linear nonholonomic
constraints  (Pfaffian constraints).
\\

\noindent{\bf Hamiltonian mechanics}.
Let $(Q,g,U)$ be a Hamiltonian mechanical system, where
$Q$ is a smooth Riemannian $n$-manifold with
a $C^\infty$ metric tensor $g$, and where
$U\in C^\infty(Q)$ denotes the potential energy.
No gyroscopic forces are taken into consideration.
Let $g^*$ denote the induced Riemannian metric on the cotangent bundle
$T^*Q$.
For a given $X\in\Gamma(TM)$, let $\theta_X$ be the 1-form defined by
$\theta_X(Y)=g(X,Y)$ for all  $Y\in\Gamma(TQ)$. It follows then that
$g(X,Y)=g^*(\theta_X,\theta_Y)$ for all  $X,Y\in\Gamma(TQ)$.

The K\"ahler metric of the previous discussion, which was denoted
by the same letter $g$, will not appear again in the sequel. Therefore,
writing $g$ for the Riemannian metric on $Q$ should hopefully not give
rise to any confusion.

In a local trivialization of
$T^*Q$, a point $x\in T^*Q$ is represented by a tuple $(q^i,p_j)$,
where $q^i$ are coordinates on $Q$,
and $p_k$ are
fibre coordinates in $T^*_q Q$, with $i,j=1,\dots,n$.
The natural symplectic 2-form associated to $T^*Q$, written in
coordinates as
$$\omega_0 = \sum_i dq^i \wedge dp_i\;=\;-d\theta_0\;, $$
is exact.
$\theta_0=p_i dq^i$ is referred to as the symplectic 1-form.

We will only consider Hamiltonians of the form
\eqn\label{hamham}H(q,p)\;=\;\frac{1}{2}\;g_q^*(p,p)\;+\;U(q)\;
      \;\;\;\in\; C^\infty(T^*Q)\;,\eeqn
that is, kinetic plus potential energy.
In local bundle coordinates, the associated Hamiltonian
vector field $X_H$ is given by
$$X_H\;=\; \sum_i \left( (\partial_{p_i}H)\partial_{q^i} -
      (\partial_{q^i}H)\partial_{p_i}\right)\; .$$
The orbits of the associated Hamiltonian flow $\Phi_t$ satisfy
\eqn\label{herodot}\dot{q}^i = \partial_{p_i} H(q,p)\;\;\;\;,\;\;\;\;
         \dot{p}_j = -\partial_{q^j}H(q,p) \;.\eeqn
The superscript dot abbreviates $\partial_t$, and will be used throughout
the discussion.

Let ${\cal A}_I$ denote the space of
smooth curves $\gamma:I\subset\R\rightarrow T^*Q$, with $I$ compact
and connected, and let $t$ denote a coordinate on $\R$.
The basis one form $dt$ defines a measure on $\R$.
The action functional is defined by
${\cal I}:{\cal A}_I\rightarrow \R$,
\eqn\label{sagittarius}{\cal I}[\gamma]&=&\int_I\;dt\;
      \left(\gamma^*\theta_0\; -\;  H\circ\gamma \right)\\
      &=&\int_I \;dt\;\left(\sum
        p_i(t)\dot{q}^i(t)-H(q(t),p(t))\right)\;,\nonumber\eeqn
with
$\dot{\gamma}=\sum(\dot{q}^i\partial_{q^i}+\dot{p}_i\partial_{p_i})$.
Denoting the base point projection by
$$\pi:T^*Q\longrightarrow Q\;,$$
let
$$c:=(\pi\circ\gamma):I\longrightarrow Q\;$$
denote the projection of $\gamma$ to $Q$.
We assume that
$\|c(I)\|$ is sufficiently small so that
solutions of (~\ref{herodot}) exist, which connect the end points
$c(\partial I)$.
Among all curves
$\gamma:I\rightarrow T^*Q$ with fixed projected
endpoints $c(\partial I)$, the ones that extremize ${\cal I}$ are
physical orbits of the system.
\\

\noindent{\bf Linear non-holonomic constraints}.
Let us next discuss the inclusion of linear
constraints on a given Hamiltonian mechanical system.
This is achieved by adding a rank $k$ distribution $W$
over $Q$ to the existing data, and by
specifying
a physical law, the H\"older variational principle,
that generates the correct physical flow on $T^*Q$.
The orbits of the
resulting dynamical system have the property that
their projections to $Q$ are $W$-horizontal.
The physics of the
H\"older principle is, for instance, discussed in \cite{Ar1}.
\\

\noindent{\bf Orthoprojectors}.
There is a $g$-symmetric tensor $\alph:TQ\rightarrow TQ$ with
$${\rm Ker}(\alph)\;=\;W^\perp\;\;\;\;,\;\;\;\;
       \alph(X)\;=\;X\;\;\;\;\forall\;X\;\in\;\Gamma(TQ)\;.$$
It will be referred to as the $g$-orthogonal projector
associated to $W$. We note that its matrix is in Mat$(n\times n,\R)$,
of rank $k$. $W^\perp$ denotes the $g$-orthogonal
complement of $W$.
The $g$-orthogonal projector associated to $W^\perp$ will be
denoted by $\bbeta$, so that
$$\alph\;+\;\bbeta\;=\;{\rm id}\;.$$

Furthermore, the dual of $W$, denoted by $W^*$,
is defined as the image of $W$ under the
isomorphism $g:TQ\rightarrow T^*Q$, and likewise for
$(W^*)^\perp := g\circ W^*$. The corresponding $g^*$-orthogonal
projectors are denoted by $\alph^\dagger$ and $\bbeta^\dagger$, respectively.
We use this notation because
the matrices of the latter are represented
by the transposed matrices of $\alph$ and $\bbeta$
in every standard coordinate chart.
We will use
the same symbols for the projectors and their matrices.
\\

\noindent{\bf Dynamics of the constrained mechanical system}.
Next, we derive the equations of motion of the constrained
mechanical system from the  H\"{o}lder  variational principle.
The use of the orthoprojectors $\alph$, $\bbeta$
is inspired by \cite{Bra}.
For a closely related approach to the Lagrangian theory of constrained
mechanical systems, cf.
\cite{CaFa}.

\begin{dfi}
A projective $W$-horizontal curve in $T^*Q$
is an embedding $\gamma:I\subset\R\rightarrow T^*Q$
whose image
$c=\pi\circ\gamma$ under the base point projection
$\pi:T^*Q\rightarrow Q$ is everywhere tangent to $W$.
\end{dfi}

Let $\gamma_s:I\rightarrow T^*Q$, with $s\in[0,1]$, be a smooth
one parameter family of curves for which the end points
$c_s(\partial I)$ are
independent of $s$ (where $c_s:=\pi\circ\gamma_s$).

\begin{dfi}
A $W$-horizontal variation of a projective $W$-horizontal curve $\gamma$
is a smooth one parameter family
$\gamma_s:\R\rightarrow T^*Q$, with $s\in[0,1]$, for which
$\frac{\partial}{\partial s}(\pi\circ\gamma_s)$ is tangent to $W$,
and $\gamma_0=\gamma$.
\end{dfi}

Let
$$\delta q^i(t):=\left.\partial_s\right|_{s=0}q^i(s,t)\;\;\;\;,\;\;\;\;
     \delta p_k(t):=\left.\partial_s\right|_{s=0}p_k(s,t)\;.$$
To any $W$-horizontal variation $\gamma_s$ of a $W$-horizontal
curve $\gamma_0$ with fixed projections of the boundaries
\eqn\label{varcond}(\pi\circ\gamma_s)(\partial I)=
      (\pi\circ\gamma_0)(\partial I)\;,\eeqn
so that $\delta q^i|_{\partial I}=0$, we associate
the action functional
$${\cal I}[\gamma_s]=\int_I \left(\sum
        p_i(s,t)\dot{q}^i(s,t)-H(q(s,t),p(s,t))\right)dt\;. $$

\begin{thm} (H\"older's principle)
A projective $W$-horizontal curve $\gamma_0:I\rightarrow T^*Q$
corresponds to a
physical motion of the constrained mechanical system if
it extremizes ${\cal I}[\gamma_s]$ among all $W$-horizontal
variations $\gamma_s$
that satisfy (~\ref{varcond}).
\end{thm}

Hence,  if
\eqn\label{varact}
     \delta{\cal I}[\gamma_s]\;=\;\sum p_i \delta q^i|_{\partial I} +\int_I
     \sum \left((\dot{p}_i-\partial_{q^i}H)\delta q^i -
     (\dot{q}^i+ \partial_{p_i}H)\delta p_i\right)\;=\;0\eeqn
for all $W$-horizontal variations of $\gamma_0$
that satisfy $\delta q^i|_{\partial I}=0$,
then $\gamma_0$ is a physical orbit.

\begin{thm}\label{dae}
In the given local bundle chart,
the Euler-Lagrange equations of the H\"older variational
principle are the differential-algebraic relations
\eqn\label{qdotcon} \dot{q} &=& \alph(q)\partial_p H(q,p)
    \\ \label{pdotcon} \alph^\dagger(q)\dot{p} &=& -\alph^\dagger(q)
       \partial_q H(q,p) \\
       \label{physleaf} \bbeta(q)\partial_p H(q,p)&=&0\;.\eeqn
\end{thm}

\prf
The boundary term
vanishes due to $\delta q^i|_{\partial I}=0$.

For any fixed value of $t$, one can write
$\delta q(t)$ as
$$\delta q(t) \;=\; \sum_{\alpha=1}^k \;f_\alpha(q(t))\; Y_\alpha(q(t))\; ,$$
where $Y_\alpha$ is a
$g$-orthonormal family of vector fields over $c(I)$ that spans $W_{c(I)}$.
Furthermore, $f_\alpha\in C^\infty(c(I))$ are
test functions obeying the boundary condition
$f_\alpha(c(\partial I))=0$.

Since $f_\alpha$ and
$\delta p$ are  arbitrary, the terms in (~\ref{varact}) that
are contracted with $\delta q$, and the ones that are contracted with
$\delta p$ vanish independently.
In case of $\delta q$, one finds
$$\int_I\; dt\;f_\alpha\;\;
      (\dot{p}+\partial_q H)_i\;Y_\alpha^i\;=0$$
for all test functions $f_\alpha$. Thus,
$(\dot{p}+\partial_q H)_i\,Y_\alpha^i=0$ for all
$\alpha=1,\dots,k$, or
equivalently, $\alph^\dagger(\dot{p}+\partial_q H)=0$,
which proves (~\ref{pdotcon}).

Since $\gamma_0$
is $W$-horizontal, $\bbeta(q)\dot{q}=0$,
so the $\delta p$-dependent term in
$\delta{\cal I}[\gamma_s]$ gives
\eqnn \int_I dt\; (\dot{q}-\alph\partial_p H)^i\;(\alph^\dagger\delta p)_i
      \;\;+\;\;\int_I dt\; (\bbeta\partial_p H)^i\;(\bbeta^\dagger\delta p)_i
      \;=\;0\;.\eeqnn
The components of $\delta p$ in the images of
$\alph^\dagger(q)$ and $\bbeta^\dagger(q)$ can be varied independently.
Thus, both terms on the second line must vanish separately,
as a consequence
of which one obtains (~\ref{qdotcon}) and (~\ref{physleaf}).
\qed

\begin{dfi}
The smooth submanifold
$$\phys\;:=\;\left\lbrace\; (q,p)
    \;\left|\;\bbeta(q)\partial_p\; H(q,p)\; =\;0\;\right\rbrace
    \;\;\subset\;\; T^*Q\right.$$
defined by (~\ref{physleaf}) will
be referred to as the physical leaf.
\end{dfi}

It contains all physical orbits of the system, that is,
all smooth path $\gamma:\R\rightarrow\phys\subset T^*Q$ that
satisfy the differential-algebraic relations of theorem {~\ref{dae}}.

\begin{thm}
Let $H$ be of the form (~\ref{hamham}). Then,
there exists a unique physical orbit $\gamma:\R^+\rightarrow\phys$ with
$\gamma(0)=x$ for every $x\in \phys$.
\end{thm}

\prf
Our strategy consists of proving that the differential
relations (~\ref{qdotcon})
$\sim$ (~\ref{physleaf}) define a unique section $X$ of $T\phys$,
where $\phys$ is regarded as an embedded submanifold of $T^*Q$.
Solution curves $\gamma:\R^+\rightarrow\phys$ of
$\dot{\gamma}(t)=X|_{\gamma(t)}$ fulfill
(~\ref{qdotcon}) $\sim$ (~\ref{physleaf})
for all initial conditions $\gamma(0)\in\phys$.
The assertion thus follows from the existence and uniqueness theorem of
ordinary differential equations.

To this end, let us cover $\phys$ with local bundle
charts of $T^*Q$ with coordinates $(q,p)$.
In case of the Hamiltonian (~\ref{hamham}), the defining relation
(~\ref{physleaf}) for $\phys$ reduces to
$$\bbeta(q)\;M^{-1}(q)\;p\;=\;M^{-1}(q)\;\bbeta^\dagger(q)\;p\;=\;0\;,$$
where $M$ denotes the matrix of $g$ ($M$ is determined by the
{\em m}ass distribution of the system),
and where one uses the $g$-orthogonality of $\bbeta$.
This implies that (~\ref{physleaf}) is equivalent to the condition
$\bbeta^\dagger(q) p=0$. Hence, $\phys$ is the common zero level
set of the $n$ component functions $(\bbeta^\dagger(q) p)_i$.

Consequently, every section
$$X\;=\;v^r(q,p)\;\partial_{q^r}\;+\;w_s(q,p)\;\partial_{p_s}$$
of $T\phys$ is annihilated by the 1-forms
$$d(\bbeta^\dagger p)_i\;=\;\partial_{q^r}(\bbeta^\dagger p)_i\;dq^r\;+\;
     \partial_{p_s}(\bbeta^\dagger p)_i \;dp_s$$
for $i=1,\dots,n$ (of which only
$n-k$ are linearly independent), on $\phys$.
This is, in vector notation, expressed by the condition
\eqnn0&=& (v^r\partial_{q^r}) \bbeta^\dagger p\; +
    (w_s\partial_{p_s})\bbeta^\dagger p
     \\ &=&
    (v^r\partial_{q^r}) \bbeta^\dagger p\;+\;\bbeta^\dagger w\;,
\eeqnn
which shows that the components $v$ of $X$
determine the projection $\bbeta^\dagger w$, so that knowing the components
$v$ and $\alph^\dagger w$ suffices to uniquely reconstruct $X$.
Consequently, the right hand sides of (~\ref{qdotcon}) and
(~\ref{pdotcon}) determine a unique section $X$ of $T\phys$, so that every
curve $\gamma:\R^+\rightarrow\phys$, with arbitrary $\gamma(0)\in\phys$,
that satisfies
$$\partial_t\gamma(t)=X(\gamma(t))\;$$
automatically fulfills (~\ref{qdotcon}) $\sim$ (~\ref{physleaf}).
This proves the assertion.
\qed

\noindent{\bf Equilibria}.
The equilibria of the constrained Hamiltonian mechanical system on
$\phys$ are obtained from the condition $\dot{q}=0$ and $\dot{p}=0$ in
(~\ref{qdotcon})
$\sim$ (~\ref{physleaf}), whereupon one arrives at
$$p=0\;\;\;\;,\;\;\;\; \alph^\dagger(q) \partial_q U(q)=0\; .$$
The critical set is thus given by
\eqn\label{pegasus}\crit_Q :=\left\lbrace q\in Q|\alph^\dagger(q)
        \partial_q U(q)=0\right\rbrace .\eeqn
An application of Sard's theorem fully analogous to the proof of
theorem {~\ref{Sardthm}} shows that generically,
this is a piecewise smooth, $n-k$-dimensional submanifold of $Q$,
(recall that the rank of
$\alph(q)$ is $k$).
\\

\noindent{\bf Symmetries}.
Let $G$ be a Lie group, and let
\eqnn\psi:G&\rightarrow& {\rm Diff}(Q) \\ h&\mapsto&\psi_h\;\;\;,\;\;\;
    \Psi_e\;=\;{\rm id}\;,\eeqnn
be a group action  on the configuration manifold so that the following
holds:

\newcounter{n1}
\begin{list}
  {(\arabic{n1})}{\usecounter{n1}\setlength{\rightmargin}{\leftmargin}}
\item Invariance of the Riemannian metrics:
      $g\circ\psi_{h}=g$ and $g^*\circ\psi_{h}=g^*$ for all $h\in G$.
\item Invariance of the potential energy:
      $U\circ\psi_{h}=U$ for all $h\in G$.
\item Invariance of the distributions:
      $\psi_{h\,*}W=W$ and $\psi_h^* W^*=W^*$ for all $h\in G$.
\end{list}

Then, we will say that the constrained Hamiltonian mechanical system
$(Q,g,U,W)$ exhibits a $G$-symmetry.

\subsection{Construction of the auxiliary extension}

Let us finally merge the nonholonomic mechanical system into a
constrained Hamiltonian systems of the type considered in the
previous sections.

To this end, we will introduce a set of generalized Dirac
constraints over the symplectic manifold $(T^*Q,\omega_0)$
in the way presented in section {~\ref{gendir}}.
They will define a suitable symplectic distribution $V$,
in a manner that the constrained Hamiltonian
system  $(T^*Q,\omega_0,H,V)$,
with $H$ given by (~\ref{hamham}), contains the constrained
mechanical system as a dynamical subsystem.
Thus, $(T^*Q,\omega_0,H,V)$
extends the mechanical system in the sense announced in the
introduction.
An early inspiration for this construction stems
from \cite{SoBr}.

We require the following properties to be satisfied by the
auxiliary constrained Hamiltonian system
$(T^*Q,\omega_0,H,V)$.

\newcounter{n2}
\begin{list}
{(\roman{n2})}{\usecounter{n2}\setlength{\rightmargin}{\leftmargin}}
\item $\phys$ is an invariant manifold under the flow $\tPhi_t$
      generated by (~\ref{eqsofmo}).
\item All orbits $\tPhi(x)$ with initial
      conditions $x\in\phys$  satisfy
      the Euler-Lagrange equations of the H\"older principle.
\item $\phys$ is marginally stable under $\tPhi_t$.
\item The critical set $\crit$ of $\tPhi_t$ is a vector
      bundle over $\crit_Q$, hence equilibria of the constrained mechanical
      system are obtained from equilibria of the extension by base point
      projection.
\item Symmetries of the constrained mechanical system extend
      to symmetries of $\tPhi_t$.
\end{list}

Let us briefly comment on  (iii) $\sim$ (v).
(iii) is of importance for numerical simulations of the
mechanical system. (iv) makes it easy to extract
information about the behaviour of the
mechanical system from solutions of the auxiliary system.
Condition (v) allows to apply reduction theory to the
auxiliary system,
in order to reduce the constrained mechanical
system by a group action, if present.
The choice for $V$ is by no means unique,
and depending on the specific problem at hand,
other conditions than (iii) $\sim$ (v)
might be more useful.
\\

\noindent{\bf Construction of $V$}.
Guided by the above requirements,
we shall now construct $V$.

To this end, we
pick a smooth, $g^*$-orthonormal family of 1-forms
$\lbrace \zeta_I\rbrace_{I=1}^{n-k}$
with
$$\zeta_I\;=\;\zeta_{Ik}(q)\;dq^k\;,$$
so that locally,
$$\langle\lbrace\zeta_1,\dots,\zeta_{n-k}\rbrace\rangle =
      \left(W^*\right)^\perp\;.$$
The defining relationship $\bbeta^\dagger(q)p=0$ for $\phys$ is equivalent
to the condition
\eqn\label{holcon}f_I(q,p):= g_q^*(p,\zeta_I(q))=0 \;\;\;
      \forall I=1,\dots,n-k\;. \eeqn
It is clear that $f_I\in C^\infty(T^*Q)$.

\newcounter{n3}
\begin{list}
  {(\Roman{n3})}{\usecounter{n3}\setlength{\rightmargin}{\leftmargin}}
\item  To satisfy conditions (i) and (iii), we require that the level surfaces
      \eqn\label{invman}
      \Mm_{\underline{\mu}}:=\left\lbrace(q,p)|f_I(q,p)=\mu_I; I=1,\dots,n-k
           \right\rbrace ,\eeqn
      with $\underline{\mu}:=(\mu_1,\dots,\mu_{n-k})$,
      are integral manifolds of $V_{r(V)}$. Here, $r(V)$ denotes
      the degree of non-holonomy of $V$, and evidently,
      $\Mm_{\underline{0}}=\phys$.

      Condition (iii) is satisfied because
      $$L(q,p)\;:=\;\sum_I\;\left|f_I(q,p)\right|^2$$
      is an integral of motion for orbits of $\tPhi_t$.
      Since $L$ grows monotonically with increasing $|\underline{\mu}|$,
      and attains its (degenerate) minimum of value zero on
      $\phys$, it is a Lyapunov function for $\phys$.
      Anything better than
      marginal stability is prohibited by energy conservation.
\item To satisfy condition (ii), we demand that  $\bbeta(q)\dot{q}=0$,
      or equivalently, that
      \eqn\label{nonhcon}\zeta_I(\dot{q})\;=\;0\;\;\;,\;\;
           \forall I=1,\dots,n-k\;,\eeqn
      shall be satisfied
      along all orbits $(q(t),p(t))$ of (~\ref{eqsofmo}),
      owing to (~\ref{qdotcon}).
\item If the constrained mechanical system exhibits a $G$-symmetry,
      characterized by a group action $\psi:G\rightarrow {\rm Diff}(Q)$
      so that $\psi_{h*}W=W\;\forall h\in G$, the local family of 1-forms
      $\lbrace\zeta_I\rbrace$ can be picked in a manner that
      $\psi_{h}^*\zeta_I=\zeta_I$ is satisfied for all
      $h\in G$ in a vicinity of the unit element $e$.
      Consequently, the functions
      $f_I(q,p)=h^*_q(\zeta_I,p)$ and their level sets
      $\Mm_{\underline{\mu}}$ are invariant under the group action.
\end{list}

The condition that (~\ref{invman})
are integral manifolds of $V_{r(V)}\supset V$ implies
that all sections of $V$ are
annihilated by the 1-forms $df_I$, for $I=1,\dots,n-k$.
Furthermore, the condition (~\ref{nonhcon}) requires $V$ to
be annihilated by the 1-forms
\eqn\label{defxi}\xi_I\;:=\;\zeta_{Ir}(q)\;dq^r\;+\;\sum_s\;0\;dp_s\;\eeqn
that are
obtained from lifting $\zeta_I$ to $T^*(T^*Q)$, with $I=1,\dots,n-k$.

\begin{prp}\label{defV}
The distribution
$$V\;:=\;\left(\bigcap_I{\rm ker}\;df_I\right)\;\;\;\bigcap\;\;\;
      \left(\bigcap_I{\rm ker}\;\xi_I\right)\;\;\subset\;T(T^*Q)$$
is symplectic.
\end{prp}

\prf
$V$ is symplectic iff its symplectic
complement $V^\perp$ is.
With the given data, the latter condition is more convenient to check.
$V^\perp$ is locally spanned by the vector fields
$(Y_1,\dots,Y_{2k})$ obtained from
\eqn\label{granite}\omega_0(Y_I,\cdot)\;=\;\xi_I(\cdot)\;\;\;,\;\;\;
      \omega(Y_{I+k},\cdot)\;=\;d f_I(\cdot) ,\eeqn
where $I=1,\dots, k$, and $\omega_0=-dp_i\wedge dq^i$.

$V^\perp$ is symplectic if and only if
$D:=\;[\omega(Y_I,Y_J)]$ has values in $GL_\R(2(n-k))$.
\\

\noindent{\bf Remark.}
In the present notation, capital indices range from
$1$ to $k$ if they label 1-forms, and from $1$ to
$2k$ if they label vector fields.
\

In local bundle coordinates,
$$\label{kabana}df_I \;=\; (\partial_{q^i}f_I)(q,p)\; dq^i \;\;+\;\;
      \zeta_{Ii}(q)\;M^{ij}(q)\;dp_j \;,$$
where $M_{ij}$ are the components of the metric tensor $g$ on $Q$,
as before.
Let us introduce the functions
$E(q):=[\zeta_{Ji}(q)]$ and $F(q,p):=[\partial_{q^j}f_K(q,p)]$, both
with values in ${\rm Mat}_\R(n\times (n-k))$,
which we use to  assemble
\eqnn K\;:=\;\left[\begin{array}{cc}
         E^\dagger&0\\F^\dagger&E^\dagger M^{-1}\end{array}\right]\;\;\;\;:\;
         T^*Q\;\longrightarrow\;{\rm Mat}_\R(2(n-k)\times 2n)\;.\eeqnn
Any component vector
$v:T^*Q\rightarrow \R^{2n}$ that locally represents an element
of $\Gamma(V)$ satisfies $K v=0$.
The symplectic structure $\omega_0$ is locally represented by
$J$, defined in (~\ref{sympJ}).
One can easily verify that the $I$-th row vector of the matrix
$K \Jj^{-1}$ is the component vector of $Y_I$.
In conclusion, introducing the matrices
\eqnn G(q)&:=&E^\dagger(q)\; M^{-1}(q)\;E(q)\\
      S(q,p)&:=& F^\dagger(q,p)\;
     M^{-1}(q)\;E(q)\;-\; E^\dagger(q)\;
     M^{-1}(q)\;F(q,p)\;  ,\eeqnn
one immediately arrives at
\eqn D& =&K\Jj K^\dagger\;=\;
     \left[\begin{array}{cc} 0& G\\-G&S
     \label{Dmat}\end{array}\right].\eeqn
Since $\zeta_I$ has been picked a $g^*$-orthonormal family
of 1-forms on $Q$, it is clear that $G(q)=\1_{n-k}$.
Thus, $D$ is invertible. This proves that
$V^\perp$ is symplectic. \qed

\noindent{\bf Construction of the projectors}.
Next, we determine the matrix  of the
$\omega_0$-orthogonal projector $\PV$, which is associated to $V$, in
the present bundle chart.
Again, it is more convenient
to carry out the construction for its complement first.

\begin{prp}
The matrix of the $\omega_0$-orthogonal projector $\PVc$
associated to $V^\perp$ (considered as a tensor field that
maps $\Gamma(T(T^*Q))$ to itself, with kernel $V$) is given by
\eqnn \PVc=\left[\begin{array}{cc}\bbeta&0\\
      T&\bbeta^\dagger\end{array}\right]\eeqnn
in the local bundle chart $(q,p)$. The matrix $T=T(q,p)$
is defined in (~\ref{coolshit}).
\end{prp}

\prf
The proof of lemma {~\ref{Pconstr}}
can be used for this proof.
The inverse of (~\ref{Dmat}) is
\eqnn D^{-1}=\left[\begin{array}{cc} S
             &-\1_{n-k}\\
             \1_{n-k}&0 \end{array}\right]\;\;,\eeqnn
where we recall that $G(q)=\1_{n-k}$.
The $I$-th column vector of the matrix
$K \Jj^{-1}$ is the component vector of $Y_I$ (we have required that
$\lbrace Y_1,\dots Y_{2(n-k)}\rbrace$ spans $V^\perp$).
This implies that $\PVc=\Jj K^\dagger D^{-1} K$.

\begin{lm}
The matrix of the $g$-orthogonal projector $\bbeta$ associated to
$W^\perp$ in the present chart is given by
\eqn\label{betaeq}\bbeta(q)\;=\;M^{-1}(q)\;E(q)\;E^\dagger(q)\;.\eeqn
\end{lm}

\prf
The construction presently carried out
for $\PVc$ can also be applied to
$\bbeta$. One only has to replace $V^\perp$ by
to $W^\perp$, and $\omega_0$ by the Riemannian
metric $g$ on $Q$.
An easy calculation immediately produces the asserted
formula.
The matrix of $\alph$ is subsequently obtained from $\alph+\bbeta=\1$.
For more details, cf. \cite{Bra}. \qed

Introducing
\eqn\label{coolshit} T(q,p)\;:=\;E(q)\;F^\dagger(q,p)\;\alph(q)\;\;-\;\;
     \alph^\dagger(q)\;F(q,p)\; E^\dagger(q)\; ,\eeqn
a straightforward calculation produces the asserted formula
for $\PVc$. \qed

\begin{cor}
In the given bundle coordinates, the matrix of $\PV$ is
\eqnn \PV=\left[\begin{array}{cc}\alph&0\\-T&\alph^\dagger
      \end{array}\right]\;, \eeqnn
where $T=T(q,p)$ is defined in (~\ref{coolshit}).
\end{cor}

\prf
This is obtained from
$\PV+\PVc =\1_{2n}$. \qed

The $\omega_0$-orthogonality of $\PV$ is represented by
$$\PV(x)\;\Jj\;=\;\Jj\;\PV^\dagger(x) $$
in the given chart.

\begin{thm}
Let $H$ be as in (~\ref{hamham}).
Then, the dynamical system
\eqn\label{salvation}\left(\begin{array}{c}\dot{q}\\\dot{p}\end{array}\right)
    =\left[\begin{array}{cc}0&\alph\\-\alph^\dagger&-T\end{array}\right]
       \left(\begin{array}{c}\partial_q H\\
       \partial_p H\end{array}\right) \eeqn
corresponding to the contrained Hamiltonian system $(T^*Q,\omega_0,H,V)$
is an extension of the constrained mechanical system $(Q,g,U,W)$.
\end{thm}

\prf
By construction, $\phys$ is an invariant manifold of the
associated flow $\tPhi_t$, hence
(~\ref{physleaf}) is fulfilled for all orbits of (~\ref{salvation})
with initial conditions in $\phys$.

The equation
$\dot{q}=\alph \partial_p H$ in (~\ref{salvation})
obviously is (~\ref{qdotcon}).

Next, using the notation
$\underline{f}:=\left(f_1,\dots,f_{n-k}\right)^\dagger$,
\eqnn\underline{f}\;=\;E^\dagger\; M^{-1}\;p\;,\eeqnn
and substituting (~\ref{coolshit}) for $T(q,p)$, the equation
for $\dot{p}$ in (~\ref{salvation}) becomes
\eqnn\dot{p}\;=\;-\alph^\dagger\; \partial_q H\;-\;E\; F^\dagger\dot{q}
     \;+\;\alph^\dagger F \underline{f}\;.\eeqnn

Since $M_{\underline{\mu}}$ are invariant manifolds
of the flow $\tPhi_t$ generated by (~\ref{salvation}),
$\partial_t f_I(q(t),p(t))$ vanishes along all orbits of (~\ref{salvation}),
so that
$F^\dagger\dot{q}+E^\dagger M^{-1}\dot{p}=0$.
This implies that
\eqn\dot{p}\;=\;-\alph^\dagger\;\partial_q H\;+\;
       E\; E^\dagger\; M^{-1}\; \dot{p}
      \;+\;\alph^\dagger\;\partial_q \left(\frac{1}{2}\underline{f}^\dagger
       \underline{f}\right)\;.\label{dotpeq1}\eeqn
Recalling that $\bbeta=M^{-1}E E^\dagger$ from
(~\ref{betaeq}), and using the fact that $\underline{f}=
\underline{0}$ on $\phys$, one  arrives at
(~\ref{pdotcon}) by multiplication with $\alph^\dagger$ from the left.
\qed

\begin{thm}
The critical set of (~\ref{salvation}) corresponds to
$$\crit\;=\;\left\lbrace (q,p)\;\left|\;q\in\crit_Q\;;\;
       p\;\in\;( W^*_q)^\perp\right\rbrace\right. \;.$$
It is a vector bundle over the base space $\crit_Q$,
with fibres given by those of  $(W^*)^\perp$.
\end{thm}

\prf
Let us start with equation (~\ref{dotpeq1}).
As has been stated above,
the second term on the right hand
side of the equality sign equals
$\bbeta^\dagger(q)\dot{p}$, and moreover, from (~\ref{holcon}), one concludes that
$$\underline{f}^\dagger
    \underline{f}\;=\;\left\|\bbeta^\dagger p\right\|_{g^*}^2\;.$$
The Hamiltonian (~\ref{hamham}) can be decomposed into
$$H(q,p)\;=\;H(q,\alph^\dagger p)\;+\;
    \frac{1}{2}\left\|\bbeta^\dagger p\right\|_{g^*}^2\;,$$
due to the $g^*$-orthogonality of $\alph^\dagger$ and $\bbeta^\dagger$,
so that (~\ref{dotpeq1}) can be written as
$$\dot{p}\;=\;-\;\alph^\dagger \partial_q H(q,\alph^\dagger p)\;+\;\bbeta^\dagger\dot{p}\;.$$

The equilibria of (~\ref{salvation}) are
therefore determined by the conditions
$$\alph^\dagger(q)p\;=\;0\;\;\;,\;\;\;\alph^\dagger(q)\;
      \partial_q H(q,\alph^\dagger p)\;=\;0\; .$$
Because $H$ depends quadratically on $\alph^\dagger p$,
the second condition
can be reduced to
$$\alph^\dagger(q)\;\partial_q U(q)\;=\;0$$
using the first condition.
Comparing this with (~\ref{pegasus}), the assertion follows.
\qed

In particular, this fact implies that every
equilibrium $(q_0,p_0)$ of the extension defines a unique
equilibrium $q_0$ on $\crit_Q$, which is simply obtained from
base point projection.
\\

\noindent{\bf Extension of symmetries}.
Let us finally prove that $(T^*Q,\omega_0,H,V)$ extends the
$G$-symmetry $\psi:G\rightarrow {\rm Diff}(Q)$
of the constrained mechanical system $(Q,g,U,W)$, if there is one.
To this end, we recall that the 1-forms $\zeta_I$ satisfy
$\psi^*_h \zeta_I$ for all $h\in G$ close to the unit.

Via its pullback, $\psi$ induces the group action
\eqnn\Psi:=\psi^*\;\;:\;\;G\times T^*Q&\longrightarrow& T^*Q\eeqnn
on  $T^*Q$.
This group action is symplectic, that is,  $\Psi_h^*\omega_0=\omega_0$
for all $h\in G$.
For a proof, consider for instance \cite{AbMa}.

The 1-forms $\xi_I$, defined in (~\ref{defxi}), satisfy
$\Psi_h^*\xi_I=\xi_I$, and likewise,
$f_I\circ\Psi_h=f_I$ is satisfied for all $h\in G$ close to the unit.
The definition
of $V$ in proposition {~\ref{defV}} thus implies that
$$\Psi_{h\,*}V\;=\;V$$
is satisfied for all $h\in G$.
Due to the fact that $\omega$ and $V$ are both $G$-invariant,
$\PV$ and $\PVc$ are also invariant under the $G$-action $\Psi$.

The Hamiltonian $H$ in (~\ref{hamham}) is $G$-invariant under $\Psi$,
by assumption on the constrained Hamiltonian
mechanical system.
Thus,  $X_H$ fulfills
$\Psi_{h*}X_H=X_H$ for all $h\in G$, which implies that
$X_H^V=\PV (X_H)$ is $G$-invariant.
\\

\noindent{\bf Stability of equilibria}.
To analyze the stability of a given equilibrium solution $q_0\in \crit_Q$,
it is necessary to
determine the spectrum of the linearization of $X_H^V$ at $x_0=(q_0,0)$.

A straightforward calculation much in the style of the discussion
above shows that in the present bundle chart,
$$DX_H^V(x_0)\;=\;\left[\begin{array}{cc}0&\alph
    M^{-1}\alph^\dagger+R^\dagger\\-\alph^\dagger D^2_{q_0}
     U\alph-R&0
   \end{array}\right](x_0)\; ,$$
where
\eqn\label{corrosion}\left[R_{jk}\right]\;:=\;
      \left[\;\partial_{q^i}U(\alph)^r_j\;(\alph)^s_k\;
      \partial_{q^s}(\alph)^i_r\;\right]\;\;\;\;
      \in\;{\rm Mat}_\R(n\times n)\; .\eeqn
Furthermore, $D^2_{q_0} U$ is the matrix of second derivatives of $U$.
The conjectural stability criterion formulated in the previous
section can now straightforwardly be applied to
$DX_H^V(x_0)$.

\subsection{\nbf The topology of the critical manifold}

Let us now discuss the global topology of the critical set of the
constrained Hamiltonian mechanical system,
defined by
$$\crit=\left\lbrace (q,p)\left|q\in\crit_Q;\alph^\dagger(q)p=0;
         \alph^\dagger(q)p=\sum_I
        \mu_I\zeta_I(q)\right\rbrace\right. .$$
Here, $\zeta_I$ is an orthonormal spanning family of one forms for the rank
$n-k$ annihilator of the rank $k$ distribution $W$, and
$$\crit_Q =\left\lbrace q\in Q|\alph^\dagger(q) \partial_q U(q)=0\right\rbrace$$
is the critical set of the physical system on $\phys$.
We recall that generically, $\crit_Q$ is a smooth $n-k$-dimensional submanifold
of $Q$. Evidently, $\crit$ is the smooth rank $n-k$ vector bundle
$$\crit=W^*_\beta\mid_{\crit_Q} $$
over the base manifold $\crit_Q$, whose fibres are given by those of the
annihilator $W_\beta^*$ of $W$.

The arguments and results demonstrated in section two can be straightforwardly
applied to the present problem.
First of all, we claim that $\crit_Q$ is normal hyperbolic
with respect to the gradient-like flow $\psi_t$ generated by
$$\partial_t q(t)=-\alph(q(t))\nabla_g U(q(t)) ,$$
and that it contains all
critical points of the Morse function $U$, but no other conditional extrema
of $U|_{\crit_Q}$ apart from those (it is gradient-like because along all of
its non-constant orbits, $\frac{d}{dt}U(t)=-g(\alph\nabla_g
U,\alph\nabla_g U)|_{q(t)}<0$ holds, as $\alph$ is an orthoprojector with
respect to the Riemannian metric $g$ on $Q$).  The fact that this is true can be
proved by substituting $M\rightarrow Q$, $H\rightarrow U$,
$P\rightarrow\alpha$, $g_{({\rm Kahler})}\rightarrow g$, and
$\crit\rightarrow
\crit_Q$ in section two, and by applying the arguments used there.
Hence, letting $\mu_i$ denote the index of the connected component
$\crit_{Qi}$, defined as the dimension of its unstable manifold, the
Conley-Zehnder inequalities (~\ref{CZ1}) imply that 
for a compact, closed $Q$, the
topological formula
\eqn\label{humdrumbum} \sum_{i,p} \lambda^{p+\mu_i} {\rm dim}H^p(\crit_{Qi}) =
     \sum_{p} \lambda^p {\rm dim}H^p(Q) +(1+\lambda) \Q(\lambda) \eeqn
holds,
where $\crit_{Qi}$ are the connected components of $\crit_Q$.  Here, $H^p$
denotes the $p$-th de Rham cohomology group with suitable coefficients, and
$\Q(t)$ is a polynomial with non-negative integer coefficients.
The argument using the Morse-Witten complexes associated to $(Q,U)$ and to
$(\crit_Q,U|_{\crit_Q})$ to derive (~\ref{humdrumbum}) can also be carried out
in the manner explained in section 2.4.
\\

Our next issue is to discuss the global topology of $\crit$. Clearly,
$\crit$ is {\em not} a compact submanifold of $T^*Q$, hence the
Conley-Zehnder inequalities of the second section, which were derived for
compact, closed, generic critical manifolds, do not apply. However,
since both $\crit$ and
$T^*Q$ are vector bundles of a particular type,
one can nevertheless prove a result that is closely
related to (~\ref{CZ1}).
Nevertheless, we claim that given the above stated properties of $\crit_Q$,
the generalized Conley-Zehnder inequalities
\eqn\label{humdrum} \sum_{i,p} \lambda^{p+\mu_i} {\rm dim}H^p_c(\crit_i) =
     \sum_{p} \lambda^p {\rm dim}H^p_c(T^*Q) +(1+\lambda) \Q(\lambda)\eeqn
are valid. In this formula, the connected components
$\crit_i$ of $\crit$ are vector bundles whose base manifolds are the connected
components
$\crit_{Qi}$ of $\crit_Q$, and the numbers $\mu_i$ are the
indices of $\crit_{Qi}$ with respect to $\psi_t$. The polynomial
$\Q(t)$ exhibits non-negative integer coefficients,
and $H^*_c$ denotes the de Rham cohomology based on differential forms with
compact supports.

In fact, (~\ref{humdrum}) is a straightforward consequence of the
circumstance that the base space of any vector bundle
is a deformation retract  the vector bundle.
Hence, $\crit_Q$, being the zero section of $\crit$,
is a deformation retract of $\crit$,
and likewise, $Q$ is a deformation retract of $T^*Q$.
Since the de Rham cohomology groups are invariant under retraction,
one infers the equality
$$H^p_c(\crit_i)\cong H^p(\crit_{Qi})\hspace{2cm}H^p_c(T^*Q)\cong H^p(Q) .$$
Hence, formula (~\ref{humdrum}) is equivalent to the assertion that
\eqn  \sum_{i,p} \lambda^{p+\mu_i} {\rm dim}H^p(\crit_{Qi}) =
      \sum_{p} \lambda^p {\rm dim}H^p(Q) +(1+\lambda) \Q(\lambda).\eeqn
But this has just been proved by
application of CZ theory to the flow $\psi_t$ on $Q$.
The weak
Conley-Zehnder inequalities derived from this result are  hence given by
$$\sum_{i} {\rm dim}H^{p-\mu_i}(\crit_{Qi}) \geq B_{p} ,$$
where $B_p$ is the $p$-th Betti number of $Q$.
In particular, for the special value $\lambda=-1$, one obtains
$$\sum_{i,p} (-1)^{p+\mu_i} {\rm dim}H^p(\crit_{Qi}) =
         \sum_{i} (-1)^{\mu_i} \chi(\crit_{Qi}) =\chi(Q) ,$$
where $\chi$ denotes the Euler characteristic.
\\

\subsection{\nit Numerically determining the critical manifold}

Let us finally formulate another possible application of the results
so far for technical purposes.

Knowledge
about the location of equilibria is very crucial for the design of
a constrained multibody system.
It is particularly desirable to know whether a chosen
set of parameters and constraints has the effect that the critical
sets are generic or not.

Technical systems in engineering applications are often very large,
so that their equilibria can usually only be constructed with
the help of numerical methods.
The analysis that we have employed in chapters two and four, aimed at
investigating global topological properties of generic critical
manifolds, inspires the following strategy.

We claim that if $U$ is a Morse function,
whose critical points are known,
and if $Q$ is compact and closed,
it is possible to numerically construct all generic
connectivity components of $\crit_Q$.

This is because generic components of
$\crit_Q$ are smooth, $n-k$-dimensional submanifolds of $Q$ containing all
critical points of $U$, and
no other critical points of $U|_{\crit_Q}$.
This information can be exploited to find sufficiently many points
on $\crit_Q$, so that suitable spline interpolation permits the
approximate reconstruction of an entire connectivity component.
To this end, one picks a vicinity of a critical point $a$ of $U$, and
uses a fixed point solver to determine neighboring zeros of
$|\alph(q)\nabla_g U(q)|^2$, which are
elements of $\crit_Q$ close to $a$.
Iterating this procedure with the critical points found in
this manner, pieces of $\crit_Q$ of arbitrary size can be
determined.

If all critical points of $U$ are a priori known,
one can proceed like this to construct all connectivity
components of $\crit_Q$ that contain critical points of $U$.
In this case,
one is guaranteed to have found all of the generic components of $\crit$ if
the numerically determined connectivity components are
closed, compact, and contain all critical points of $U$.

We remark that the determination of the critical points of
a Morse function $U:Q\rightarrow\R$
is a difficult numerical
task. Attempting to find critical points by simulating the
gradient flow generated by $-\nabla_g U$ is presumably
very time costly, because the critical points define a
thin set in $M$.
Their existence, however, is ensured by the topology of $Q$
if the latter is nontrivial,
as a result from the Morse inequalities.

Another remark is that all critical points $a$
at which ${\rm Jac}_a(\alph \nabla_g U)$
has a reduced rank, are elements of the non-generic
part of $\crit_Q$.
If there are such exceptional critical points in a technically relevant
region of $Q$, they can be removed by a small
local modification of the system parameters and the constraints.
\\

\pagebreak


\begin{thebibliography}{99}

\bibitem{AbMa} R. Abraham, J. E. Marsden, 'Foundations of mechanics',
Benjamin/Cummings, (1978).

\bibitem{Ar} V. I. Arnol'd, 'Mathematical methods of classical mechanics',
Second Edition, Graduate Texts in Mathematics {\bf 60}, Springer Verlag (1989).

\bibitem{Ar1} V. I. Arnol'd,  'Dynamical systems III', Encyclopedia of
Mathematics {\bf 3}, Springer Verlag (1988).

\bibitem{AuBr} D. M. Austin, P. J. Braam, 'Morse-Bott theory and equivariant
cohomology', in 'The Floer memorial volume', eds. H. Hofer, C. H. Taubes, A.
Weinstein, E. Zehnder, Birkh\"auser Verlag (1995).

\bibitem{BeRi} A. Bellaiche, J.-J. Risler (eds.), 'Sub-Riemannian geometry',
Birkh\"auser Verlag (1996).

\bibitem{Bi} J. - M. Bismut, 'The Witten complex and the degenerate Morse
inequalities', J. Diff. Geom., {\bf 23}, 207 - 240 (1986).

\bibitem{BlKrMaMu} A.M. Bloch, P.S. Krisnaprasad, J.E. Marsden, R.M. Murray,
'Nonholonomic mechanical systems with symmetry', Arch. Rational Mech. Anal.,
{\bf 136}, 21-99, (1996).

\bibitem{Bo} R. Bott, 'Nondegenerate critical manifolds', Ann. Math. {\bf 60},
No. 2, 248 - 261, (1954).

\bibitem{Bo1} R. Bott, 'Morse theory indomitable', Publications
math\'ematiques, {\bf 68}, 99 - 114 (1989).

\bibitem{Bo2} R. Bott, 'Lectures on characteristic classes and
foliatons', in R. Bott, S. Gitler, I.M. James, 'Lectures on algebraic
topology',  Lecture notes in mathematics, {\bf 279}, Springer Verlag (1972).

\bibitem{Bra} H. Brauchli, 'Mass-orthogonal formulation of equations of motion
for multibody systems', J. Appl. Math. Phys. (ZAMP), {\bf 42} , 169 - 182
(1991).

\bibitem{Bra2} H. Brauchli, 'Efficient description and geometrical
interpretation of the dynamics of constrained systems', Computational Methods
in Mechanical Systems '97, J. Angeles, E. Zakhariev (eds.), Springer Verlag
Berlin Heidelberg (1998).

\bibitem{Bry} R. L. Bryant, 'Lectures on symplectic geometry', in 'Geometry
and quantum field theory', D. S. Freed, K. K. Uhlenbeck (eds.), IAS/College
Park mathematics series Vol. I (1995).

\bibitem{CaFa} F. Cardin, M. Favretti, 'On nonholonomic and vakonomic dynamics
of mechanical systems with nonintegrable constraints', J. Geom. Phys., {\bf
18}, 295 - 325 (1996).

\bibitem{CoZe} C. Conley, E. Zehnder, 'Morse type index theory for flows and
periodic solutions of Hamiltonian equations', Comm. Pure Appl. Math., {\bf
37}, 207 - 253 (1984).

\bibitem{DuFoNo} B. A. Dubrovin, A. T. Fomenko, S. P. Novikov, 'Modern geometry
- methods and applications', Vol. III, Springer Verlag, (1985).

\bibitem{Fl} A. Floer, 'Witten's complex and infinite dimensional Morse
theory', J. Diff. Geom., {\bf 30}, 207 - 221 (1989).

\bibitem{Ge} Z. Ge, 'Betti numbers, characteristic classes and sub-Riemannian
geometry', Illinois J. Math., {\bf 36}, No. 3, 372 - 403 (1992).

\bibitem{Gr2} M. Gromov, 'Carnot - Caratheodory spaces seen from within', in
'Sub-Riemannian geometry', eds. A. Bellaiche, J.-J. Risler, Birkh\"auser
Verlag, (1996).

\bibitem{Hi} M. W. Hirsch, 'Differential topology', Springer Verlag New York,
(1976).

\bibitem{HoZe} H. Hofer, E. Zehnder, 'Symplectic invariants and Hamiltonian
dynamics', Birkh\"auser Verlag, (1994).

\bibitem{Jo} J. Jost, ' Riemannian geometry and geometric analysis', Springer
Verlag Berlin Heidelberg, (1995).

\bibitem{KoMa} W. S. Koon, J. E. Marsden,
'The Hamiltonian and Lagrangian approaches to the dynamics of nonholonomic
systems', Rep. Math. Phys., {\bf 40}, 21-62 (1997).

\bibitem{MaRa} J. E. Marsden, T. Ratiu, 'Introduction to mechanics and
symmetry', Springer Verlag New York (1994).

\bibitem{McSa} D. McDuff, D. Salamon, 'Introduction to symplectic topology',
Clarendon Press (1995).

\bibitem{Mi} J. Milnor, 'Morse theory', Princeton Univ. Press, Princeton, N.
J. (1963).

\bibitem{Mi2} J. Milnor, 'Topology from the differentiable viewpoint',
Princeton Landmarks in Mathematics, Princeton University Press (1997).

\bibitem{Sw} M. Schwarz, 'Morse Homology', Birkh\"auser Verlag (1993).

\bibitem{Sm} S. Smale, 'Morse inequalities for a dynamical system', Bull.
Amer. Math. Soc., {\bf 66}, 43 - 49 (1960).

\bibitem{Sm1} S. Smale, 'On gradient dynamical systems', Ann. Math., {\bf
74}, No. 1, 199 - 206 (1961).

\bibitem{SoBr} M. Sofer, O. Melliger, H. Brauchli, 'Numerical behaviour of
different formulations for multibody dynamics', Numerical Methods in
Engineering '92, Ch. Hirsch et al (eds.), Elsevier Science Publishers,
Amsterdam (1992).

\bibitem{Sp} E. Spanier, 'Algebraic Topology', Springer Verlag New York (1966).

\bibitem{Str} R. Strichartz, 'Sub-Riemannian geometry', J. Diff.
Geom., {\bf 24}, 221 - 261 (1986).

\bibitem{vdSMa} J. Van der Schaft, B.M. Maschke, 'On the Hamiltonian formulation
of nonholonomic mechanical systems', Rep. Math. Phys., {\bf 34}, 225-233 (1994).

\bibitem{Wi} E. Witten, 'Supersymmetry and Morse theory', J. Diff. Geom. {\bf
17}, 661 - 692, (1982).

\bibitem{We} R. W. Weber, 'Hamiltonian systems with constraints and their
meaning in mechanics', Arch. Rat. Mech. Anal. {\bf 91}, 309 - 335 (1986).

\bibitem{YoKa} H. Yoshimura, T. Kawase, 'A duality principle in nonholonomic mechanical
systems', to appear in Meccanica, (2000).

\bibitem{Ze} E. Zehnder, 'The Arnold conjecture for fixed points of symplectic
mappings and periodic solutions of Hamiltonian systems', Proceedings of the
International Congress of Mathematicians, Berkeley, California, USA (1986).

\bibitem{ZeBlMa} D. V. Zenkov, A. M. Bloch, J. E. Marsden, 'The
energy-momentum method for the stability of nonholonomic systems',
Dyn. Stab. of Systems, {\bf 13}, 123-166 (1998).

\end{thebibliography}
\end{document}